\documentclass[superscriptaddress,groupedaddress,nofootnoteinbib,12pt,english]{article}
\pdfoutput=1
\usepackage[latin9]{inputenc}
\usepackage[letterpaper]{geometry}
\usepackage{amsmath}
\usepackage{amssymb}
\usepackage{cite}
\usepackage{babel}
\usepackage{jcappub}
\usepackage{amssymb}
\usepackage{amsmath}
\usepackage{amstext}
\usepackage{graphicx,epsfig}
\usepackage{epsfig}
\usepackage{verbatim}
\usepackage{fancyhdr}
\usepackage{fancybox}
\usepackage{color}
\usepackage{ulem,bbold}
\usepackage{enumitem}
\usepackage{subfigure}
\usepackage{bbm}
\usepackage{parskip}

\newcommand{\Comment}[1]{{}}
\definecolor{MyDarkBlue}{rgb}{0.15,0.15,0.45}
\usepackage[linktocpage=true]{hyperref}
\hypersetup{
colorlinks=true,
citecolor=MyDarkBlue,
linkcolor=MyDarkBlue,
urlcolor=MyDarkBlue,
}

\newcommand{\be}{\begin{equation}}
\newcommand{\ee}{\end{equation}}
\newcommand{\bea}{\begin{eqnarray}}
\newcommand{\eea}{\end{eqnarray}}
\newcommand{\nn}{\nonumber}

\def\({\left(}
\def\){\right)}

\newcommand{\half}{\frac{1}{2}}

\newcommand{\bR}{ \bar{R}}
\newcommand{\e}{\bar{e}}
\newcommand{\bg}{\bar g}

\newcommand{\bnabla}{\bar\nabla}
\newcommand{\om}{ \bar{\omega}}

\newcommand{\calD}{{\cal D}}
\newcommand{\calbD}{\bar {\cal D}}

\def\ba{\begin{eqnarray}}
\def\ea{\end{eqnarray}}

\def\L{\mathcal{L}}

\def\nn{\nonumber}

\def\d{\mathrm{d}}
\def\mn{_{\mu \nu}}

\def\({\left(}
\def\){\right)}
\def\ie{{\it i.e. }}

\def\mpl{M_{\rm Pl}}
\def\p{\partial}
\def\stu{St\"uckelberg }

\numberwithin{equation}{section}

\begin{document}


$\phantom{.}$\vspace{-40pt}

\title{Evidence for and Obstructions to Non-Linear Partially Massless Gravity\vspace{-20pt}}
\author{Claudia de Rham${}^{a,}$\footnote{E-mail address: \Comment{\href{mailto:claudia.derham@case.edu}}{\tt claudia.derham@case.edu}},
Kurt Hinterbichler${}^{b,}$\footnote{E-mail address: \Comment{\href{mailto:khinterbichler@perimeterinstitute.ca}}{\tt khinterbichler@perimeterinstitute.ca}},
Rachel A. Rosen${}^{c,}$\footnote{E-mail address: \Comment{\href{mailto:rar2172@columbia.edu}}{\tt rar2172@columbia.edu}}
and Andrew J.~Tolley${}^{a,}$\footnote{E-mail address: \Comment{\href{mailto:andrew.j.tolley@case.edu}}{\tt andrew.j.tolley@case.edu}}\vspace{-30pt}}
\affiliation{${}^{a}$Department of Physics, Case Western Reserve University, 10900 Euclid Ave, Cleveland, OH 44106, USA\\
${}^{b}$Perimeter Institute for Theoretical Physics, 31 Caroline St. N., Waterloo, ON, N2L 2Y5, Canada\\
${}^{c}$Physics Department and Institute for Strings, Cosmology, and Astroparticle Physics, Columbia University, New York, NY 10027, USA}

\abstract{Non-linear partially massless (PM) gravity, if it exists, is a theory of massive gravity in which the graviton has four propagating degrees of freedom. In PM gravity, a scalar gauge symmetry removes one of the five modes of the massive graviton.  This symmetry ties the value of the cosmological constant to the mass of the graviton, which in turn can be kept small in a technically natural way.  Thus PM gravity could offer a compelling solution to the old cosmological constant problem.  In this work we look for such a theory among the known ghost-free massive gravity models with a de Sitter reference metric.  We find that despite the existence of strong supporting evidence for the existence of a PM theory of gravity, technical obstructions arise which preclude its formulation using the standard massive gravity framework.}

\maketitle

\setcounter{footnote}{0}

\newpage

\section{Introduction}
\parskip=5pt
\normalsize

One motivation for the recent resurgence of interest in massive gravity has been the need to explain the observed cosmic acceleration.  Massive gravity, as an alternative to general relativity at cosmological scales, could provide the means to address the old cosmological constant problem (why the cosmological constant is zero) as well as an alternative explanation of the new dark energy problem (see \cite{Hinterbichler:2011tt} for a review).  Here we investigate the possibility of a `Partially Massless' theory of gravity which may offer a new approach to the old cosmological constant problem.

A generic theory of massive gravity requires a fiducial or reference metric upon which the theory of gravitons is defined.  Maximally symmetric reference metrics are of special importance, as the notion of a graviton can then be defined via the irreducible representations of the associated isometry group.  In Minkowski space-time, the representation theory of the Poincar\'{e} group tells us that a massive spin-two field has five degrees of freedom.  Generically, massive spin-two representations of the de Sitter (dS) group also have five degrees of freedom.  However, for dS there is also an exceptional case.  For a specific value of the graviton mass relative to the dS curvature, representation theory tells us that a massive graviton has only four propagating degrees of freedom.  The field theory that describes this particle is known to exist at the linear level and is referred to as `Partially Massless' (PM) gravity  \cite{Deser:1983mm,Higuchi:1986py,Deser:2001pe,Deser:2001us,Deser:2001wx,Deser:2001xr,Deser:2004ji,Zinoviev:2001dt}.

In the linear PM theory there is a scalar gauge symmetry which renders the helicity-0 mode pure gauge and hence unphysical.  If PM gravity exists as a non-linear theory (whether this is possible is the topic of this article), there will exist a single gauge symmetry which eliminates the helicity-0 mode at all orders.  As with diffeomorphism invariance in general relativity, the PM gauge symmetry should be sufficient to fix the low energy form of the theory.  This is in contrast to generic massive gravity, for which no known symmetry enforces the form of the interactions.  Thus, PM gravity would have many of the aesthetic virtues of Einstein's theory despite being fundamentally an infrared modified theory of gravity.

However, PM gravity differs from both general relativity and generic massive gravity in one crucial aspect.  In PM gravity the value of the cosmological constant is not free, but is tied to the mass of the graviton via the gauge symmetry.  To be precise, in $D$ space-time dimensions, the cosmological constant $\Lambda$ is given in terms of the graviton mass $m$ by $\Lambda = (D-1) m^2/2$. The existence of a gauge symmetry ensures (in the absence of anomalies) that quantum corrections preserve this relation. This means that the old cosmological constant problem, which arises due to the large sensitivity of the value of $\Lambda$ to quantum corrections, is replaced with the significantly more manageable problem of looking at quantum corrections to the mass of the graviton.  The latter problem is under much better control due to the fact that in the limit $m \rightarrow 0$ PM gravity reduces to general relativity which has an additional three gauge symmetries. The existence of these gauge symmetries in the massless limit is sufficient, via the 't Hooft naturalness argument, to protect the value of the graviton mass from large quantum corrections \cite{ArkaniHamed:2002sp,deRham:2012ew}. The implication is that, if PM gravity exists as a non-linear theory, then it can provide a technically natural solution to the old cosmological constant problem which is based in symmetry and does not rely on degravitation or screening \cite{ArkaniHamed:2002fu,Dvali:2007kt}.

PM gravity, if it exists, has a number of additional virtues over generic massive gravity that are a consequence of the absence of the helicity zero mode. There is no vDVZ discontinuity \cite{vanDam:1970vg,Zakharov:1970cC} in the limit $m \rightarrow 0$ provided that we maintain the PM relations between the mass and cosmological constant.  In this theory there is no need for the Vainshtein mechanism \cite{Vainshtein:1972sx} -- the extra scalar can't cause any fifth force problem because it simply doesn't exist -- and hence no associated strong coupling.
The theory would have a higher cutoff  $\Lambda_2\sim (\mpl m)^{1/2}$ than that of generic massive gravity $\Lambda_3\sim \(\mpl m^2\)^{1/3}$, as the worst non-renormalizable operators are associated with the now absent scalar field couplings.  Finally, the absence of a helicity zero mode could potentially remove any issues to do with superluminalities \cite{deRham:2011pt,Burrage:2011cr,Izumi:2013poa,Deser:2013eua}.  This is because superluminalities typically arise when the Galileon-like derivative couplings of the scalar are expanded around non-trivial backgrounds.  In the PM theory these couplings would be absent.  Combined with the potential of PM gravity to solve the old cosmological constant problem, these reasons compel us to look for the existence of the non-linear PM theory.  (Ref. \cite{Deser:2013uy}, which appeared as we were writing, asks this same existence question, and reaches the same conclusion we do, using different methods.)

To avoid any confusion, it is worth pointing out that PM gravity is different from the minimal model of dRGT (de Rham, Gabadadze, Tolley \cite{deRham:2010iK,deRham:2010kj}), corresponding to a specific choice of parameters of the mass term with Minkowski as the reference for which no interactions are present in the decoupling limit keeping the scale $\Lambda_3$ fixed. In the minimal model, the helicity-0 mode of the graviton is still present, but one ought to probe slightly higher energy scales to see its first interactions. In PM gravity on the other hand, the helicity-0 mode is fully absent.

Furthermore, PM gravity is also different from recently found FRW solutions of massive gravity where the kinetic term disappears. In these solutions the fundamental theory does have a helicity-0 mode but the latter cancels on a specific background, signaling a strongly coupled issue, \cite{Gumrukcuoglu:2011zh,D'Amico:2012pi,Wyman:2012iw,Khosravi:2013axa}.
We also emphasize that if it exists, PM gravity is different from others models of Lorentz-violating massive gravity for which the helicity-0 mode is also absent, \cite{Rubakov:2008nh}. Even though there is not a Lorentz symmetry about dS, it is still a maximally symmetric spacetime with the same amount of symmetry as Minkowski.

\bigskip
In the rest of this section we review the linear partially massless theory.   We also review the ghost-free dRGT theories of massive gravity among which we search for a non-linear completion of PM gravity.  We present a compact formulation of our candidate non-linear PM action, as was originally determined by the analysis of \cite{deRham:2012kf}.
We emphasize however that the analysis of \cite{deRham:2012kf} was restricted to the decoupling limit. In \cite{deRham:2012kf} the symmetry was extended non-linearly in the gauge parameter but not in the field and only gave a hint onto how the symmetry could get generalized fully non-linearly if it did exist. However as will be shown in this manuscript, this symmetry does not exist non-linearly.

In Section \ref{evidencesection} we give independent evidence in support of our candidate action being the unique potentially partially massless theory.   In particular, we perform an analysis for the vector modes in the decoupling limit, an analysis away from the decoupling limit using an FRW ansatz for the dynamical metric and its perturbations, and a brute force perturbative analysis of the non-linear PM symmetry up to cubic order in the fields.  We find that the helicity-zero mode is indeed absent for the unique choice of the dRGT coefficients in the PM candidate theory.  In particular, we give a detailed derivation of the following (known) result: while at quadratic order in fields the PM theory exists in arbitrary space-time dimensions, at cubic order it exists only for $D=4$ \cite{Zinoviev:2006im,Joung:2012hz}.  We can also determine the non-linear gauge symmetry of the partially massless theory, in the FRW ansatz and to cubic order in interactions for a general metric.

In Section \ref{obstructionsection} we show that, despite the promising evidence of the previous section, the candidate partially massless theory does not have the required gauge invariance to all orders.  We see the first indication of this by looking at anisotropic cosmologies.   We then push the perturbative analysis of the non-linear PM symmetry up to quartic order in the fields, making no assumption as to the form of the potential for the graviton.  We show that at quartic order no PM symmetry exists without the introduction of either non-canonical derivative terms or additional fields.  These conclusions agree with those of \cite{Zinoviev:2006im,Deser:2013uy}.  The Lagrangian that most nearly realizes the PM symmetry at the non-linear level coincides with an expansion of the dRGT ghost-free Lagrangian with parameters given by the decoupling limit of \cite{deRham:2012kf}, i.e., our candidate theory.  Nevertheless, away from the decoupling limit an obstruction unavoidably remains.  We conclude with a discussion of possible ways out and future directions for partially massless gravity.

\bigskip
{\bf Conventions}:   We use the mostly plus metric signature convention, $\eta_{\mu \nu} = (-,+,+,+,\ldots)$ and we work in arbitrary $D\geq 3$ space-time dimensions, unless otherwise stated.

\subsection{The linear PM theory\label{linearsection}}

Consider the Fierz-Pauli theory of a massive graviton $h_{\mu\nu}$ propagating on a maximally symmetric background $\bg_{\mu\nu}$,
\bea \nn {\cal S}&=&\int d^Dx\ \sqrt{-\bg}\left[ -{1\over 2}\bnabla_\alpha h_{\mu\nu}\, \bnabla^\alpha h^{\mu\nu}+\bnabla_\alpha h_{\mu\nu} \,\bnabla^\nu h^{\mu\alpha}-\bnabla_\mu h\,\bnabla_\nu h^{\mu\nu}+\half \bnabla_\mu h\bnabla^\mu h\right. \\ &&\left. +{\bR\over D}\left( h^{\mu\nu}h_{\mu\nu}-\half h^2\right)-\frac{1}{2}m^2(h_{\mu\nu}h^{\mu\nu}-h^2)\right]. \label{curvedmassivelin}\eea
Here $m^2$ is the graviton mass.  The metric, covariant derivatives and constant curvature $\bR$ are those of the background.  For most choices of $m^2$ this action propagates the usual number of degrees of freedom of a massive graviton (e.g., five for $D=4$).  For $m^2=0$ there is linearized diffeomorphism symmetry $ \delta h_{\mu\nu}=\bnabla_\mu\xi_\nu+\bnabla_\nu\xi_\mu,$
and the action propagates the ${}$ degrees of freedom of a massless graviton (e.g., two for $D=4$).

A clean way to see the degrees of freedom is to introduce, following \cite{ArkaniHamed:2002sp}, the \stu fields $A_\mu$ and $\phi$ through the replacement
\be \label{astukcurve} h_{\mu\nu}\rightarrow h_{\mu\nu}+\bnabla_\mu A_\nu+\bnabla_\nu A_\mu+2\,\bnabla_\mu\bnabla_\nu\phi.\ee
There are now two new gauge symmetries
\bea & \delta h_{\mu\nu}=\bnabla_\mu \xi_\nu+\bnabla_\nu \xi_\mu,\ \ \delta A_\mu=-\xi_\mu, \\
& \delta A_{\mu}=\bnabla_\mu \Lambda,\ \ \ \delta\phi=-\Lambda, \eea
and fixing the gauge $A_\mu=0,\ \phi=0$ recovers the original action (\ref{curvedmassivelin}).

We next make the following conformal transformation which serves to unmix the scalar and the tensor
\be \label{curvedconfo} h_{\mu\nu}= h^\prime_{\mu\nu}+{2\over D-2}m^2\phi\, \bg_{\mu\nu}.\ee
The resulting action is
\bea \nn {\cal S}=\int d^Dx\ \mathcal{L}_{m=0}(h^\prime)&+&\sqrt{-\bg}\bigg[-\frac{1}{2}m^2(h^\prime_{\mu\nu}h^{\prime\mu\nu}-h^{\prime 2})-\half m^2\, F_{\mu\nu}F^{\mu\nu}+{2\over D}m^2\bR\, A^{\mu}A_{\mu} \\
 \nn &-&2m^2 \left(h^\prime_{\mu\nu}\bnabla^\mu A^\nu-h^\prime\bnabla_\mu A^\mu\right)+2m^2\({D-1\over D-2}m^2-{\bR\over D}\)\left(2\phi\,\bnabla_\mu A^\mu+h'\phi\right)\\ \label{massivenolimit}
&-&2m^2 \({D-1\over D-2}m^2-{\bR\over D}\)\left((\partial\phi)^2-m^2 {D\over D-2}\phi^2\right) \bigg] ,\eea
where $\mathcal{L}_{m=0}(h^\prime)$ is the Lagrangian for a massless graviton and $F_{\mu\nu}\equiv\partial_\mu A_\nu-\partial_\nu A_\mu$.

Here we see that for the special value
\be \label{partialmasslessvalue} \bR={D(D-1)\over D-2}m^2,\ee
the dependence on $\phi$ completely cancels out of (\ref{massivenolimit}).  Tracing back through the \stu replacements and field redefinitions, we conclude that the original Lagrangian \eqref{curvedmassivelin} has the gauge symmetry
\be \label{partialmasslesssym} \delta h_{\mu\nu}=\(\bnabla_\mu\bnabla_\nu+{m^2\over D-2}\, \bg_{\mu\nu}\)\alpha,\ee
where $\alpha$ is a scalar gauge parameter.  This symmetry looks like a tuned combination of a linear diffeomorphism with diff parameter $\sim\partial_\mu\alpha$, and a linearized Weyl transformation with parameter $\sim\alpha$.  This is the partially massless theory
\cite{Deser:1983mm,Higuchi:1986py,Deser:2001pe,Deser:2001us,Deser:2001wx,Deser:2001xr,Deser:2004ji,Zinoviev:2001dt}.  Due to the enhanced gauge symmetry (\ref{partialmasslesssym}), this theory propagates one fewer degree of freedom than the generic massive graviton, so in $D=4$ it carries four degrees of freedom rather than five.

To more clearly see what is happening for various values of $m^2$ vs. $\bR$, we take the high energy limit
\be m^2\rightarrow 0,\ \ \ \bR\rightarrow 0,\ \ \ {m^2\over \bR}\ \ {\rm fixed},\label{limitlin}\ee
and keep fixed the canonically normalized fields ${\hat A_\mu}={m A_\mu}, \ \hat\phi=m^2\phi,$ so as to preserve the number of degrees of freedom in the limit.  What remains is a decoupled action for a massless graviton, massless vector and massless scalar,
\be \nn {\cal S}=\int d^Dx\ \mathcal{L}_{m=0}(h^\prime)+\sqrt{-\bg}\bigg[-\half m^2F_{\mu\nu}F^{\mu\nu}-2m^2 \({D-1\over D-2}m^2-{\bR\over D}\)(\partial\phi)^2 \bigg]
\ee

Looking at the signs of the various kinetic terms, we see that the vector field is ghost-like unstable\footnote{Note that there is no equivalent of the Breitenlohner-Freedman bound \cite{Breitenlohner:1982bm} for massive gravitons in AdS, since we have a ghost-like instability just as soon as $m^2<0$.} for $m^2<0$.  At $m^2=0$, where the  sign is turning over, only the tensor propagates and the action \eqref{curvedmassivelin} has the enhanced linearized diffeomorphism symmetry of linearized general relativity $ \delta h_{\mu\nu}=\bnabla_\mu \xi_\nu+\bnabla_\nu \xi_\mu$.

\begin{figure}[h!]
\begin{center}
\epsfig{file=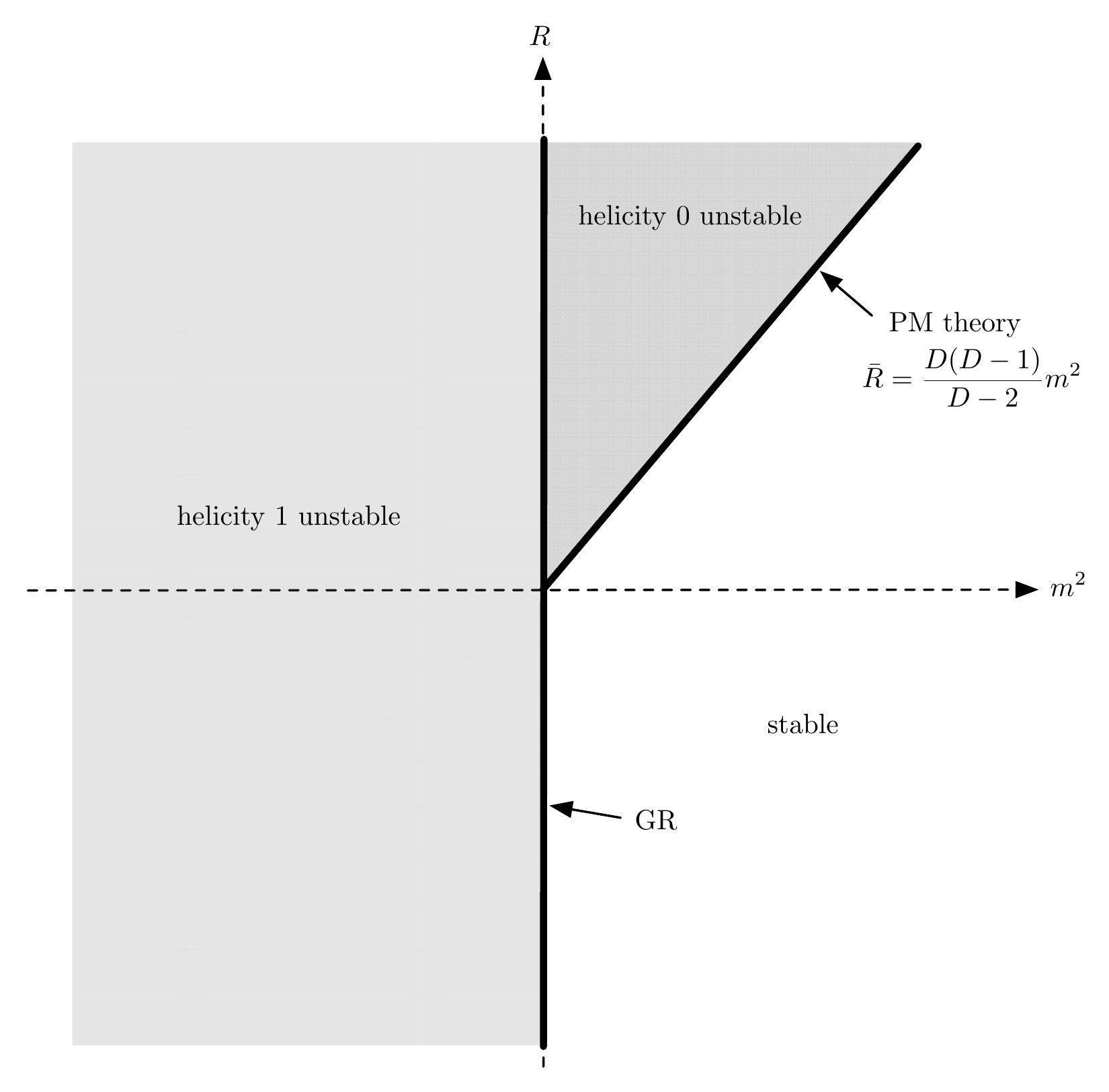,height=4in,width=4.0in}
\caption{Regions of stability for massive gravity on maximally symmetric backgrounds.}
\label{plot1}
\end{center}
\end{figure}

When $m>0$, the vector is stable but the scalar is unstable unless $m^2>{D-2\over D(D-1)}\bR$.  This is the Higuchi bound \cite{Higuchi:1986py}, and when it is saturated the scalar kinetic term disappears and we have the enhanced scalar gauge symmetry \eqref{partialmasslesssym} of the partially massless theory.  Note that because of the stability requirement $m^2>0$, the partially massless theory is only stable on dS space $\bR>0$ (which happens to be the correct sign of the cosmological constant in our universe).  In flat space, $\bR=0$, nothing like it exists, and the novel gauge symmetry \eqref{partialmasslesssym} merges with the diffeomorphisms.  See figure \ref{plot1} for a summary of the various regions in the $\bR,m^2$ plane.

The partially massless theory propagates a particle which lives in an exotic irreducible representation of the dS group\footnote{This particle can also be given an electromagnetic-like interpretation \cite{Deser:2006zx}, complete with duality \cite{Deser:2013xb}.} \cite{Deser:1983mm,Higuchi:1986py,Deser:2001pe,Deser:2001us,Deser:2001wx,Deser:2001xr,Deser:2004ji,Deser:2006zx,Zinoviev:2001dt}.  There is no flat space counterpart to this representation -- in the flat space limit the particle breaks up into a massless graviton and a massless photon.  They are only seen to be unified on large distance scales comparable to Hubble.

The partially massless symmetry ties the value of the graviton mass to the curvature of the background dS space-time, through the relation \eqref{partialmasslessvalue}.  Said another way, a bare cosmological constant is forbidden by the partially massless symmetry.  Indeed, at the linear level, a bare cosmological constant is $\sqrt{-g}\sim h$, and this is not invariant under the symmetry \eqref{partialmasslesssym}, $\delta h\not=0$.  In contrast, a large bare cosmological constant is allowed in GR because the cosmological constant term $h$ is invariant under linearized diffeomorphisms.   At least at the linear level then, the partially massless symmetry is a rare example of a symmetry which can rule out an arbitrary cosmological constant.

\subsection{The candidate non-linear theory}
Generically, adding higher order interaction terms to the known linear PM Lagrangian will destroy the partially massless symmetry and there will be more propagating modes at the non-linear level.  The question, then, is whether it is possible to have a fully non-linear theory of partially massless gravity for a particular choice of interaction terms, which has a scalar gauge symmetry like the linear theory.

The recent developments in massive gravity allow us to revisit this question.  The dRGT theory of massive gravity \cite{deRham:2010iK,deRham:2010kj} provides a non-linear completion of the Fierz-Pauli action that is free of the Boulware-Deser ghost \cite{Boulware:1973my} at the fully non-linear level \cite{Hassan:2011hr,deRham:2011rn,deRham:2011qq,Hassan:2011tf,Hassan:2011ea,Mirbabayi:2011aa,Golovnev:2011aa,Hassan:2012qv,Hinterbichler:2012cn,Kluson:2012wf}.  The original ghost-free dRGT model was proposed with a Minkowski reference metric \cite{deRham:2010iK,deRham:2010kj}, but this was subsequently generalized to a generic reference metric \cite{Hassan:2011tf,Hassan:2011zd}.  Just like dRGT massive gravity, the PM theory should be free of the Boulware-Deser (BD) ghost instability.  Since the dRGT models are the unique ghost-free massive gravity theories in D dimensions (at least with the conventional Einstein-Hilbert kinetic term, we'll say more about this later), then we expect that the PM gravity theory will be a special case of the dRGT massive gravity models formulated with a dS reference metric.

We start by looking for a non-linear PM theory within the class of ghost-free dRGT Lagrangians of massive gravity, formulated with respect to a non-dynamical dS reference metric $\bg_{\mu\nu}$,
\be
\label{Lbeta}
{\cal L} =\frac{\mpl^{D-2}}{2}\sqrt{-g} \left(R[g]-2\Lambda-\frac{m^2}{4} \sum_{n=0}^D \beta_n S_n(\sqrt{g^{-1}\bg}) \right) \, .
\ee
Here the dynamical metric is given by $g_{\mu\nu}$.  The symmetric polynomials $S_n(M)$ of a matrix $M^\mu_{\ \nu}$ are
\be S_n(M)=M^{ [\mu_1}_{\ \mu_1}\cdots M^{ \mu_n]}_{\ \mu_n}, \ee
with $S_0\equiv 1$.  The curvature $\bR$, cosmological constant $\Lambda$ and Hubble constant $H$ of the reference metric $\bar g_{\mu\nu}$ are related by
\be \Lambda=\left(D-2\over 2D\right)\bR={(D-1)(D-2)\over 2}H^2.\label{backgroundrelations}\ee

Not all the $\beta$'s are independent: to ensure that $g_{\mu\nu}=\bar g_{\mu\nu}$ is a solution we enforce tadpole cancellation which gives
\be
\label{tadpole}
D! \ \sum_{k=0}^D \frac{\beta_k}{k!(D-k)!} = (D-1)! \ \sum_{k=1}^D \frac{\beta_k}{(k-1)!(D-k)!} \,.
\ee
(For $D=4$, this yields $\beta_0 = -\left(3\beta_1 +3\beta_2+\beta_3\right)$.)  We may absorb one further coefficient by taking $m^2$ to be the mass squared of the graviton fluctuations around $\bar g_{\mu\nu}$, which is enforced by the relation
\be
-D! \ \sum_{k=0}^D \frac{\beta_k}{k!(D-k)!} +(D-2)! \ \sum_{k=2}^D \frac{\beta_k}{(k-2)!(D-k)!} = -8\, .
\ee
(For $D=4$, this yields $\beta_1+2\beta_2+\beta_3 = 8$.)  Finally, since $S_{D}(M) =\det M$ the term in \eqref{Lbeta} proportional $\beta_N$ is a constant and can be ignored, so we set $\beta_N=0$.  In total, the theory has $D-2$ free parameters, in addition to the mass and cosmological constant.

We will see in the next section that various lines of evidence (including the earlier analysis of \cite{deRham:2012kf}) point to a dRGT theory with one particular choice of coefficients which has the potential to propagate a partially massless graviton non-linearly.  These coefficients are
\be
\label{PMalpha}
\beta_0=-4(D-1),\ \ \beta_2={8\over D-2},
\ee
with all other $\beta$'s zero.
In addition, the mass $m^2$ should be at the partially massless value \eqref{partialmasslessvalue}.  Plugging these into (\ref{Lbeta}), and using the relations \eqref{backgroundrelations} for the background, we have our candidate $D$-dimensional PM Lagrangian:
\be
\label{LPM}
{\cal L}_{\rm PM}=\frac{\mpl^{D-2}}{2} \sqrt{-g} \left(R[g]-2H^2 \,S_2(\sqrt{g^{-1}\bg}) \right) \, . \\
\ee
Other than the value of the dS radius, there are no free parameters left in the theory.

We note that in $D=4$, we have the property $ \sqrt{-g} \,S_2(\sqrt{g^{-1}\bg}) = \sqrt{-\bg} \,S_2(\sqrt{\bg^{-1}g}) $ \cite{Hassan:2011zd}.  Thus if we promote the PM candidate Lagrangian to a bimetric Lagrangian by introducing an Einstein-Hilbert kinetic term for $\bg_{\mu\nu}$, this new Lagrangian will enjoy a $Z_2$ symmetry under the interchange of $g_{\mu\nu}\leftrightarrow \bg_{\mu\nu}$.  The values of the $\beta$'s presented here do not coincide with those found in \cite{Hassan:2012gz,Hassan:2012rq} (where the authors looked for a partially massless theory in a bimetric setup, using different criteria than those used here), that is, simply plugging in a fixed de-Sitter for the second metric in the candidate bi-metric theory does not reproduce the candidate single metric theory.  
However, the theories do in fact agree if the massive gravity limit of the candidate bi-metric theory is taken consistently \cite{Fasiello:2013woa} -- one scales the coefficients such that the degrees of freedom of the second graviton are retained but rendered free and decoupled from the partially massless graviton, whose action then reproduces \eqref{LPM}.

\subsection{PM action in vielbein form}

The non-linear PM action \eqref{LPM} can be expressed compactly using vielbeins and differential forms\footnote{A frame-like formulation of partially massless gravity at linear level was developed in \cite{Skvortsov:2006at}.}.  We introduce a vielbein 1-form $e^a$ and curvature 2-form $R^{ab}[e]$ for the dynamical metric $g_{\mu\nu}$, as well as a vielbein 1-form $\e^a$ and curvature 2-form $\bR^{ab}[\e] =H^2 \, \e^a \wedge \e^b $ for the dS reference metric $\bg_{\mu\nu}$.  We invoke the arguments of \cite{Hinterbichler:2012cn}, which allow us to make the following replacement,
\be
\int d^D x \sqrt{-g} \,S_2(\sqrt{g^{-1}\bg}) ~~\rightarrow~~ \frac{1}{2!(D-2)!} \int \epsilon_{a_1 \ldots a_D}\, \e^{a_1} \wedge \e^{a_2} \wedge e^{a_3} \wedge \ldots \wedge e^{a_D} \, .
\ee
With this notation, the PM candidate action given by (\ref{LPM}) can be written in terms of the difference of the dynamical curvature and the dS reference curvature,
\be
\label{SPM}
{\cal S}_{PM} = \frac{\mpl^{D-2}}{2!(D-2)!} \int  \epsilon_{a_1 \ldots a_D} (R^{a_1 a_2}-\bR^{a_1a_2})\wedge e^{a_3} \wedge \ldots \wedge e^{a_D} \, .
\ee

Let us define the relative spin connection as $\Omega^{ab} \equiv \omega^{ab}-\om^{ab}$.  Then there are many equivalent ways of rewriting the above expression.  In particular, we have
\bea
R^{ab}-\bR^{ab}&=& \calD \Omega^{ab}-\Omega^a_{~c} \wedge \Omega^{cb} \\
&=& \calbD \Omega^{ab}+\Omega^a_{~c} \wedge \Omega^{cb} \\
&=& \tfrac{1}{2} (\calD+\calbD)\Omega^{ab} \, .
\eea
We use $\calD$ and $\calbD$ to denote the derivatives which are covariant under local Lorentz transformations, for the dynamical metric and reference metric, respectively.

In this language it is straightforward to demonstrate the existence of the usual linear PM symmetry.  At lowest order in fields, the above action is invariant under,
\be
\label{vtransf}
{\delta e^a} =  \calbD\((\e^{-1})^{a \mu}\partial_\mu \alpha\)+H^2 \e^a \alpha \, ,
\ee
with gauge parameter $\alpha$.

\section{The Evidence\label{evidencesection}}
\label{sec:Evidence}

In this section, we present evidence that the candidate ghost-free dRGT Lagrangian has a partially massless symmetry non-linearly.

\subsection{The decoupling limit\label{decoupling}}

In section \ref{linearsection}, we saw at the linear level through the \stu analysis how the longitudinal mode $\phi$ disappears from the action at the special value of the mass \eqref{partialmasslessvalue}.
Starting with the Lagrangians \eqref{Lbeta}, one can perform an analogous analysis at the full non-linear level.  The \stu replacement \eqref{astukcurve} must be supplemented by the appropriate non-linear terms \cite{ArkaniHamed:2002sp}.  Once this is done, the theory is an effective field theory which looks like \eqref{massivenolimit} at linear level, but now has interaction terms suppressed by various energy scales.  The lowest such scale (We'll work in $D=4$ in this section for ease of displaying the scales) is $\Lambda_3\sim (\mpl m^2)^{1/3} \sim (\mpl H^2)^{1/3}$.

The action can then be greatly simplified by taking the high-energy decoupling limit, generalizing \eqref{limitlin}
\be m\rightarrow 0,\ \ \bR\rightarrow 0,\ \ M_P\rightarrow\infty,\ \ \ \ \ {m^2\over \bR},\Lambda_3 \ \ \text{fixed}.\ee
The decoupling limit focuses in on the leading non-linearities, leaving them intact while sending all the others to zero.
If the theory propagates a PM graviton non-linearly and has only four degrees of freedom, then the $\phi$ mode should disappear from the action completely (perhaps after some field redefinitions, generalizing the conformal transformation \eqref{curvedconfo} at linear level).  A necessary condition for this is that the $\phi$ mode disappear from the decoupling limit Lagrangian.

This analysis was done in \cite{deRham:2012kf} for arbitrary $D$, and the result is that there is a unique choice of the $\beta_n$ such that the helicity-zero
 mode of the massive graviton is absent in the decoupling limit.  These coefficients are precisely \eqref{PMalpha}, leading to the candidate Lagrangian \eqref{LPM}.

The analysis of  \cite{deRham:2012kf} considered only the tensor and scalar modes of the effective theory.  This is a consistent truncation, but if the partially massless theory exists, the scalars should also disappear from the vector-scalar interaction which are present at the scale $\Lambda_3$, and should in fact disappear completely from the effective theory (possibly after some field re-definitions).  In this subsection, we will assume that this is the case, and ask what it implies for the resulting effective theory.

If the scalar has completely disappeared, then the effective field theory only contains tensor and vector modes, and the lowest interaction scale that can appear in this theory is the one carried by the vector self-interactions, which carries the scale
$\Lambda_2\sim\(M_P m\)^{1/2} \sim\(M_P H\)^{1/2}$.  The interaction terms carry one derivative per $A$, and so they take the form
$\L_A^{(n)}\sim \frac{(\partial \hat A)^n}{\Lambda_{2}^{2n-4}}\,,$
in terms of the canonically normalized vector field $\hat A_\mu=\Lambda_{2} A_\mu$.

Since there cannot be interactions at energy scales below $\Lambda_{2}$ in a PM theory of gravity, one should be able to consider the following limit
\be\label{vectordecoupling}
\mpl\rightarrow \infty,\hspace{15pt} m, H\rightarrow 0,\hspace{20pt}{\rm keeping}\hspace{20pt} \ {m\over H}, \Lambda_{2}\ \ \rm{fixed}\,.
\ee
In this limit, one has a flat-space theory propagating a free linearized helicity-2 mode and a self-interacting vector degree of freedom, which decouples from the graviton, so this corresponds to a genuine decoupling limit of PM gravity. The decoupled vector Lagrangian is symbolically of the form
\be
\label{fself}
\L_{A,  {\rm dec}}\sim \hat F_{\mu\nu}^2+\frac{(\partial \hat A)^3}{ \Lambda_{2}^{2}}+\frac{(\partial \hat A)^4}
{\Lambda_{2}^4}+\cdots\,.
\ee

At linear level, this Lagrangian propagates only two degrees of freedom since $F\mn^2$ respects the Maxwell $U(1)$ gauge symmetry.  Non-linearly, the vector field decoupling limit Lagrangian should continue to propagate only two DOF. This implies that the interactions should remain gauge invariant under some gauge symmetry which starts as the Maxwell symmetry and then possibly gets corrections at higher order in $A$,
\be
\delta A_\mu=\partial_\mu\Lambda+\cdots\, .
\ee
To see whether such a symmetry can ever exist non-linearly, let us start with a massive graviton $g_{\mu\nu}$ on an (A)dS background $\bar g_{\mu\nu}$. To make the argument as general as possible, we work in what follows perturbatively with an arbitrary mass term (although we will see soon how the mass term is constrained to the coefficients of \eqref{LPM}, at least up to cubic order in perturbations).

Thus our starting point is a Lagrangian containing the Einstein-Hilbert kinetic term with cosmological constant $\Lambda$, and an arbitrary mass term
\bea
{\cal L}&&={\cal L}_{\rm EH}+{\cal L}_{\rm m} \nn\\
&&=\frac{\mpl^{D-2}}{2} \left[ \sqrt{-g}\(R[g]-2\Lambda\)-{m^2\over 4}\sqrt{-\bar g}\left( [h^2]-[h]^2+c_1[h^3]+c_2[h^2][h]+c_3[h]^3\right.\right. \nn\\
&&\left.\left. +d_1[ h^4]+d_2 [ h^3] [ h] +d_3[ h^2 ]^2 +d_4[ h^2][ h]^2+d_5[ h]^4 +\cdots\right)\right]\,, \label{startlag}
\eea
where the arbitrary mass term is expressed in powers of $h_{\mu\nu}=g_{\mu\nu}-\bar g_{\mu\nu}$, and indices on $h$ are raised and lowered with $\bar g_{\mu\nu}$.  The mass term is the Fierz-Pauli term at quadratic order, followed at higher order by every possible contraction of $h_{\mu\nu}$ with an arbitrary coefficient in front of each.

To derive the contributions to the Lagrangian \eqref{fself} in the decoupling limit we make the replacement \cite{ArkaniHamed:2002sp}
\be\label{vectorstuka}
h_{\mu\nu}\rightarrow \partial_\mu A_\nu+\partial_\nu A_\mu-\partial_\mu A^\rho\partial_\nu A_\rho,
\ee
in the mass term,
and take the metric and all covariant derivatives to be flat.  We can do this because by hypothesis all the couplings to $h$ disappear in the decoupling limit \eqref{vectordecoupling}, as do any further corrections due to the (A)dS curvature of the reference metric.

We expand the Lagrangian \eqref{fself} in powers of $A$,
\be {1\over \mpl^{2}m^2}{\cal L}_A={\cal L}_{(2)}+ {\cal L}_{(3)}+{\cal L}_{(4)}+\cdots,\ee
and look for a scalar gauge transformation with parameter $\alpha$, in powers of $A$,
\be \delta A_\mu=\hat{L}_\mu\alpha,\ \ \ \ \hat{L}_\mu=\hat{L}^{(0)}_\mu+\hat{L}^{(1)}_\mu+\hat{L}^{(2)}_\mu+\cdots.\ee
If the Lagrangian is gauge invariant, we should have the Bianchi identity
\be\label{Abianchi}
\hat{L}_{\mu}{\delta {\cal L}_A\over \delta A_\mu}=0\,.
\ee

At lowest order, quadratic in $A$, we have
${\cal L}_{(2)}=-\frac 18 F_{\mu\nu}^2$,
which is invariant under
$\hat{L}^{(0)}_\mu \Lambda=\partial_\mu\Lambda$.

At cubic order in $A$, the Lagrangian \eqref{startlag} with the replacement \eqref{vectorstuka} gives
\ba
\label{cubicLag}
{\cal L}_{(3)}&=&-\frac 14 \Big[ (-2+3c_1)\,
 \partial_\nu A_{\mu}\, \partial_\rho A^\nu \,\partial^\rho A^{\mu}+c_1\,  \partial_\nu A_{\mu}\, \partial^\mu A^{\rho}\,\partial_\rho A^\nu+4c_3 \,(\partial\cdot A)^3\\
&&\phantom{-\frac 14 \Big[}+2 c_2 \,(\partial\cdot A) \,\partial^\nu A^{\mu}\, \partial_\mu A_{\nu}+2(1+c_2)\,(\partial\cdot A)\,\partial^\mu A^{\nu}\,\partial_\mu A_{\nu}
\Big]\,.\nn
\ea
Next, we write the most general transformation $\hat{L}_{\mu}^{(1)}$, first order in $A$, with up to three derivatives (though only up to two actually appear, because we can't contract three), with an arbitrary coefficient in front of each term,
\bea
\hat{L}_{\mu}^{(1)}\alpha=&& C_1\, A_\mu\alpha+C_2\,  A_\mu\square\alpha+C_3\,  A^\nu \partial_\mu\partial_\nu\alpha+C_4\, \partial_\mu A^\nu\partial_\nu\alpha \nn\\
&&+C_5\, \partial^\nu A_\mu\, \partial_\nu\alpha+C_6\,  (\partial \cdot A)\, \partial_\mu\alpha+C_7\, \partial_\mu\partial_\nu A^\nu\, \alpha+C_8\, \square A_\mu\alpha\,.
\eea

At third order in $A$, the identity \eqref{Abianchi} reads
\be
\label{cubicAcond}
\hat{L}_{\mu}^{(0)}{\delta {\cal L}_{(3)}\over \delta A_\mu}+\hat{L}_{\mu}^{(1)}{\delta {\cal L}_{(2)}\over \delta A_\mu}=0\,.
\ee
Demanding this hold fixes the cubic mass term to the values,
\be\label{CAvalues}
c_1=1,\ c_2=-{5\over 4},\ c_3={1\over 4}\,,
\ee
which correctly matches the cubic mass terms of our candidate theory \eqref{LPM}.   The gauge transformation is fixed to be
\be \hat{L}_{\mu}^{(1)}\alpha={1\over 2}F_{\mu\nu}\partial^\nu\alpha+ \partial_\mu\left[ C_3 A_\nu\, \partial^\nu\alpha +C_6 (\partial\cdot A) \alpha\right].
\ee
The unfixed coefficient $C_6$ corresponds to the freedom to redefine the gauge parameter $\alpha \rightarrow \sim \(\partial\cdot A\)\, \alpha$ and
$C_3$ corresponds to the freedom to redefine the gauge parameter $\alpha \rightarrow \sim A_\nu\partial^\nu\alpha $.

While we displayed the results of this subsection for $D=4$, this analysis does not depend on the number of dimensions $D$, and the coefficients \eqref{CAvalues} are the same for any $D$, and match those of \eqref{LPM} for any $D$.

In $D=4$, the effective field theory of PM gravity at distances shorter than Hubble is a theory of a massless tensor and a vector.  If we take the graviton mass to be the observed value of Hubble today, and the Planck mass to be what is observed, then the cutoff scale comes out to
\be \Lambda_2\sim \ {\rm neutrino\ mass\ scale}.\ee
This cutoff is parametrically higher than the cutoff $\Lambda_3$ generically present in dRGT massive gravity, because the worst offending scalar interactions are absent.
It's worth noting that this is the sort of effective theory one would obtain by integrating out all massive particles from the Standard Model; the only massless degrees of freedom are the photon and graviton, and the cutoff is at the neutrino scale, where the first massive particle comes in.

\subsection{Non-linear gauge symmetry in mini-superspace\label{frwsection1}}

The candidate action \eqref{LPM} propagates a partially massless mode at linear level, the scalar mode is absent in the scalar tensor sector of the $\Lambda_3$ decoupling limit, and it has the required gauge symmetry at cubic order in the $\Lambda_2$ decoupling limit.  The key question we would like to answer is whether or not the PM symmetry exists in the full theory, with all non-linearities and beyond the decoupling limit.  While this is a difficult question to address at the fully non-linear level for an arbitrary metric, we can simplify the analysis by adopting a particular ansatz for the dynamical metric and checking for the existence of the PM symmetry there.

Here we study the full non-linear theory in the simplified case of mini-superspace, i.e. for an FRW ansatz.  Consider the PM Lagrangian in the form (\ref{LPM}), only with an arbitrary coefficient $\lambda$ in front of the mass term, so that we can see what happens as $\lambda\rightarrow H^2$,
\be
{\cal S}={\mpl^{D-2} \over 2} \int \d^Dx\ \sqrt{-g}\left(R[g]-2\lambda S_2(\sqrt{g^{-1}\bar g})\right) \, .
\ee
We plug in a flat FRW ansatz in the flat slicing for the reference dS metric, with a dynamical scale factor $a(t)$ and lapse $N(t)$ (note that there is no diffeomorphism symmetry to set the lapse to unity),
\be
\label{FRW_metric}
g_{\mu\nu}=-N(t)^2\d t^2+a(t)^2\d\vec x^2,\ \ \ \bar g_{\mu\nu}=-\d t^2+e^{2Ht}\d\vec x^2 \, .
\ee
Up to a total derivative, the action becomes
\be
{\cal S}= \mpl^{D-2}  \int \d t\ \left[ - {(D - 1)(D - 2)\over
   2} {a^{D - 3} \dot a^2\over N}-2 \lambda {D - 1\over 2} a^{D - 3} e^{H t}\(a + {D - 2\over 2} N e^{H t}\) \right]\, .
\ee
 The lapse $N(t)$ appears algebraically, so we may eliminate it through its equation of motion,
 \be
 \label{Nsol}
 N={1\over \sqrt{\lambda}}e^{-Ht}\dot a \, .
 \ee
Plugging back in the Lagrangian, we find up to a  total derivative,
\be
{\cal S}= \mpl^{D-2}  \int \d^4x\ (D-1) \sqrt{\lambda}  e^{H t} a^{D-2} (H - \sqrt{\lambda}) \,.
\ee
For the generic case $\lambda\not=H^2$, the $a$ equation of motion tells us $a=0$ and the Lagrangian is inconsistent (which is an instance of dRGT massive gravity having no flat FRW solutions \cite{D'Amico:2011jj}).  But for the special case
\be
\lambda=H^2 ,
\ee
the Lagrangian becomes empty, meaning there is a gauge symmetry with gauge parameter $\epsilon(t)$ that allows us to set the scale factor to anything we please,
\be\label{FRWgaugex}
\delta a=\epsilon,\ \ \ \ \delta N={1\over H}e^{-Ht}\dot \epsilon \,,
\ee
In this theory, cosmology is pure gauge: \textit{any} arbitrary function is a solution for the scale factor, with the lapse determined by \eqref{Nsol}.  The transformation \eqref{FRWgaugex} should be the mini-superspace form of the full non-linear PM gauge symmetry.

In terms of the metric, this implies that the PM Lagrangian is invariant under the following transformation,
\be
\delta g_{\mu\nu}={\rm diag}(-2N{1\over H}e^{-Ht}\dot \epsilon, 2a\epsilon,\cdots) \, .
\ee
The transformation here is only first order in derivatives.  If we perform the change of variables $\epsilon = \tfrac{1}{2}e^{Ht}(-H\dot{\alpha}+H^2\alpha)$ we find that this transformation is entirely consistent with the linear PM symmetry (\ref{partialmasslesssym}) with gauge parameter $\alpha(t)$.

\subsection{Perturbations in FRW}

Following \cite{Fasiello:2012rw}, let us consider cosmological perturbations around FRW for dRGT massive gravity on dS.   Requiring the coefficient of the kinetic term of the helicity-zero
 mode to be positive is what gives rise to the generalization of the Higuchi bound \cite{Higuchi:1986py}. Specializing to four dimensions $D=4$, this was derived in \cite{Fasiello:2012rw} to be (we convert the $\alpha_n$ coefficients used there to the $\beta_n$ coefficients used here)
\be
m^2(H) =\frac{m^2}{8} \frac{H}{H_0} \left( \beta_1+2 \beta_2\frac{H}{H_0}+ \beta_3 \frac{H^2}{H_0^2}\right) \ge 2 H^2\, ,
\ee
where now $H$ is the Hubble parameter of the dynamical metric, and $H_0$ that for the reference metric. The dynamical quantity $m^2(H)$ is the actual mass of the graviton for the fluctuations when the dynamic metric and the background metric are not the same.

If the PM symmetry exists then the kinetic term for the helicity-zero
 mode should vanish altogether. This corresponds to saturating the bound. However the bound must be saturated for all values of $H$ since this bound is applicable to any spatially flat FRW geometry  \cite{Fasiello:2012rw}. The only way this is possible is if the coefficient of each order of the polynomial in $H$ balances on each side of the equation. That is in order to satisfy
\be
m^2(H) = \frac{m^2}{8} \frac{H}{H_0} \left( \beta_1+2\beta_2\frac{H}{H_0}+ \beta_3\frac{H^2}{H_0^2}\right) = 2 H^2
\ee
for all $H$ we must satisfy
\ba
&&\beta_1=0  \\
&&\beta_2=8 H_0^2/m^2 \\
&& \beta_3=0.
\ea
Using the fact that the PM condition requires $m^2 = 2 H_0^2$, we find $\beta_2 =4$.  Along with the tadpole cancellation condition (\ref{tadpole}), this gives $\beta_0=-12$.  These are precisely the same coefficients as those of our candidate theory \eqref{PMalpha} in $D=4$.   I.e., they are the same coefficients as the those found in the decoupling limit analysis of \cite{deRham:2012kf} (expressed here in terms of $\beta_n$).

In other words, adopting the usual Higuchi bound condition $m^2=2 H_0^2$, along with the same coefficients \eqref{PMalpha} of the dRGT model found in the decoupling limit analysis, guarantees that the kinetic term for the helicity-zero mode vanishes automatically for any FRW geometry.  That the cancellation takes place for this specific choice of parameters demonstrates a nontrivial consistency check.  This is an independent confirmation of the PM candidate Lagrangian, performed away from the decoupling limit.

\subsection{The full theory at cubic order\label{fullcubicss}}

In this subsection we will perform a brute force systematic analysis, away from the decoupling limit and for an arbitrary dynamical metric, so as to determine order by order in non-linearity whether or not a PM theory of gravity exists.  Our starting point is a Lagrangian containing the Einstein-Hilbert kinetic term with cosmological constant $\Lambda$, and an arbitrary mass term written in powers of $h_{\mu\nu}=g_{\mu\nu}-\bar g_{\mu\nu}$.  Indices on $h_{\mu\nu}$ are raised and lowered with the reference metric $\bar g_{\mu\nu}$,
\ba
{\cal L}&=&{\cal L}_{\rm EH}+{\cal L}_{\rm m} \nn\\
&=&\frac{\mpl^{D-2}}{2} \Big[ \sqrt{-g}\(R[g]-2\Lambda\)-{m^2\over 4}\sqrt{-\bar g}\({\cal L}^{(2)}_{\rm m}
+{\cal L}^{(3)}_{\rm m}+{\cal L}^{(4)}_{\rm m}+\cdots\)\Big] \, ,
\ea
with
\ba
{\cal L}^{(2)}_{\rm m}&=&b_1 [h^2]+b_2[h]^2,\\
{\cal L}^{(3)}_{\rm m}&=&c_1[h^3]+c_2[h^2][h]+c_3[h]^3,\\
{\cal L}^{(4)}_{\rm m}&=&d_1[h^4]+d_2 [ h^3] [ h] +d_3[ h^2 ]^2 +d_4[ h^2][ h]^2+d_5[ h]^4\,,
\\ && \vdots \nn
\ea
Here we do not restrict ourselves to the dRGT interactions -- the mass term is simply every possible contraction of $h_{\mu\nu}$ with an arbitrary coefficient in front of each.
Again, the curvature of the reference metric  $\bR$ is related to the cosmological constant $\Lambda$ by \eqref{backgroundrelations}.

If a PM theory of gravity exists, it must have a scalar gauge symmetry, and be invariant under a transformation of the form
\be
\delta h_{\mu\nu}={\hat L}_{\mu\nu}\alpha \, ,
\ee
where ${\hat L}_{\mu\nu}$ is some operator and $\alpha$ is the gauge parameter.  Gauge invariance then gives us the Bianchi identity
\be
\label{gaugeinv}
{\hat L}_{\mu\nu}{\delta{\cal L}\over \delta h_{\mu\nu}}={\hat L}_{\mu\nu}\left[-\half\sqrt{-g}\(G^{\mu\nu}+\Lambda g^{\mu\nu}\)+{\delta{\cal L}_{\rm m}\over \delta h_{\mu\nu}}\right]=0 \, .
\ee
The operator ${\hat L}$ can be expanded in powers of $h$,
\be
{\hat L}={\hat L}^{(0)}+{\hat L}^{(1)}+\cdots \, .
\ee
In what follows we will attempt to determine the coefficients of the mass term and of the terms appearing in the perturbative expansion of ${\hat L}$, in order to solve (\ref{gaugeinv}) order-by-order\footnote{The is the unitary gauge version of a similar analysis done in Ref.~\cite{Zinoviev:2006im}, and we find equivalent conclusions.}.

\subsubsection{Quadratic order}

At lowest order, the statement for gauge invariance \eqref{gaugeinv} reads
\be
\label{2ndordercond}
{\hat L}^{(0)}_{\mu\nu}\left[-\half\sqrt{-g}\left.\(G^{\mu\nu}+\Lambda g^{\mu\nu}\)\right|_{(1)}+{\delta{\cal L}^{(2)}_{\rm m}\over \delta h_{\mu\nu}}\right]=0 \, ,
\ee
where here and in what follows, the subscript $|_{(n)}$ designates the expansion to $n^{\rm th}$ order in $h\mn$.
We restrict ourselves to gauge symmetries with at most two derivatives.  The most general transformation which does not involve any power of the metric perturbation then has three possible terms:
\be
\label{transf_0}
{\hat L}^{(0)}_{\mu\nu}\alpha=B_1\,  \bnabla_\mu\bnabla_\nu\alpha+B_2\, \bg_{\mu\nu}\alpha+B_3 \, \bg_{\mu\nu}\bar\square\alpha \, ,
\ee
where $\bnabla$ represents the covariant derivative with respect to the background dS metric, and we have put an arbitrary coefficient in front of each possible term.

Plugging the transformation \eqref{transf_0} into \eqref{2ndordercond}, we find that the coefficients of the mass terms and of the transformation law get fixed. To see this explicitly we can first fix coefficients starting from the highest derivatives: requiring terms of the form $\bnabla\bnabla\bnabla\bnabla h$ in \eqref{2ndordercond} to vanish fixes $B_3=0$.  Then we work our way down, requiring terms of the form $\bnabla\bnabla h$ to vanish, and then zero derivative terms $h$.  This procedure fixes
\be
b_1=-b_2={2\Lambda\over D-1},\ \ \ B_2={2\Lambda\over (D-1)(D-2)}B_1,\ \ \ B_3=0 \,.
\ee
(Notice that there is also a solution with $b_1=b_2=B_2=B_3=0$, which just corresponds to massless GR with the scalar part of diffeomorphism invariance.) Scaling $B_1=1$ by absorbing it into the gauge parameter, we find the `semi-conformal' transformation of \eqref{transf_0}:
\be
\delta h_{\mu\nu}^{(0)}= \bnabla_\mu\bnabla_\nu\alpha+{2\Lambda\over (D-1)(D-2)}\, \bg_{\mu\nu}\alpha \, .
\label{zeroortrans}
\ee

We find that the symmetry fixes the graviton mass to the partially massless value
\be
\label{m2value}
m^2={2\Lambda\over D-1} \, .
\ee
Finally, the symmetry fixes the Fierz-Pauli structure: $b_1=-b_2=1$.  The is a sign that the partially massless symmetry knows about the absence of ghosts.

\subsubsection{Cubic order}

At cubic order, the statement for gauge invariance \eqref{gaugeinv} reads
\be
\label{3ndordercond}
{\hat L}^{(0)}_{\mu\nu}\left[-\half\sqrt{-g}\left.\(G^{\mu\nu}+\Lambda g^{\mu\nu}\)\right|_{(2)}+{\delta{\cal L}^{(3)}_{\rm m}\over \delta h_{\mu\nu}}\right]+{\hat L}^{(1)}_{\mu\nu}\left[-\half\sqrt{-g}\left.\(G^{\mu\nu}+\Lambda g^{\mu\nu}\)\right|_{(1)}+{\delta{\cal L}^{(2)}_{\rm m}\over \delta h_{\mu\nu}}\right]=0 \, .
\ee

For ${\hat L}^{(1)}$, the most general form up to two derivatives has 18 terms (again all indices moved with $\bar g_{\mu\nu}$), which we write with arbitrary coefficients:
\bea &&{\hat L}^{(1)}_{\mu\nu}\alpha= \nn\\
&& C_1\, h_{\mu\nu}\bar\square\alpha+C_2\, h_{(\mu}^{\ \ \lambda}\bnabla_{\nu)}\bnabla_\lambda\alpha+C_3\, h\bnabla_\mu\bnabla_\nu\alpha \nn\\
&& +C_4\, \bnabla_{(\mu}h_{\nu)}^{\ \ \lambda}\bnabla_\lambda\alpha+C_5\, \bnabla_{(\mu}h\bnabla_{\nu)}\alpha+C_6\, \bnabla_{\lambda}h_{\mu\nu}\bnabla^\lambda\alpha+C_7\, \bnabla_\lambda h_{(\mu}^{\ \ \lambda}\nabla_{\nu)}\alpha\nn \\
&& +C_8\, \bnabla_\mu\bnabla_\nu h\, \alpha+C_9\, \bnabla_\lambda \bnabla_{(\mu}h_{\nu)}^{\ \ \lambda}\,\alpha+C_{10}\, \bar\square h_{\mu\nu}\,\alpha \nn\\
&& +C_{11}\,\bg_{\mu\nu}h\bar\square\alpha+C_{12}\,\bg_{\mu\nu}h^{\lambda\sigma}\bnabla_\lambda\bnabla_\sigma\alpha \nn\\
&&+C_{13}\,\bg_{\mu\nu}\bnabla_\sigma h^{ \sigma\lambda}\bnabla_\lambda\alpha+C_{14}\,\bg_{\mu\nu}\bnabla_\lambda h\bnabla^\lambda\alpha \nn\\
&& +C_{15}\,\bg_{\mu\nu}\bar\square h\, \alpha+C_{16}\,\bg_{\mu\nu}\bnabla_{\lambda}\bnabla_\sigma h^{\lambda\sigma}\,\alpha \nn\\
&& +C_{17}\,h_{\mu\nu}\alpha+C_{18}\,\bg_{\mu\nu}h\alpha. \label{generalcubict}
\eea
These are organized as follows:  We start with terms containing two derivatives.  First are the terms which are not proportional to $\bg_{\mu\nu}$.  The first line has the terms with both derivatives on $\alpha$, the second line has the terms with one derivative on $\alpha$, and the third line has the terms with no derivatives on $\alpha$.  Next we write the terms proportional to $\bg_{\mu\nu}$.  The fourth line has the terms with both derivatives on $\alpha$, the fifth line has the terms with one derivative on $\alpha$, and the sixth line has the terms with no derivatives on $\alpha$.

We substitute this general expression into \eqref{3ndordercond} and fix the coefficients $C_i$ for the gauge transformation as well as the coefficients $c_i$ in the mass term by
demanding that the equality \eqref{3ndordercond} hold.  Requiring the various terms of the form $h \bnabla\bnabla\bnabla\bnabla h$ (6 possible contractions) to vanish fixes $C_1=C_{11}=C_{12}=0$.  Then we work our way down, requiring terms of the form $\bnabla h\bnabla\bnabla\bnabla h$ to vanish (11 possible contractions), then terms of the form $\bnabla\bnabla h\bnabla\bnabla h$ (13 possible contractions), then the two derivative terms $h\bnabla\bnabla h$ (5 possible contractions) and $\bnabla h\bnabla h$ (5 possible contractions), and finally the zero derivative terms $h^2$ (2 different contractions), fixing coefficients along the way.  We have 21 free coefficients to kill these 42 terms, so we need 21 miracles.

The cubic mass term coefficients are fixed to the values
\bea {\cal L}=\half \left[ \sqrt{-g}\(R[g]-2\Lambda\)-{m^2\over 4}\sqrt{-\bar g}\left( [h^2]-[h]^2-[h^3]+{5\over 4}[h^2][h]-{1\over 4}[h]^3+\cdots\right)\right],\nn\\
\eea
with $m^2$ fixed to \eqref{m2value}.  The coefficients of the mass term determined by this procedure agree with those found in an expansion of the PM candidate Lagrangian (\ref{LPM}).

The gauge transformation is fixed to be
\bea \delta_{(1)}h_{\mu\nu}=&& \half h_{(\mu}^{\ \ \lambda}\bnabla_{\nu)}\bnabla_\lambda\alpha-\half \bnabla_{(\mu}h_{\nu)\lambda}\bnabla^\lambda\alpha+\half\bnabla_\lambda h_{\mu\nu}\bnabla^\lambda\alpha-{\Lambda\over 2}{D-6\over (D-1)(D-2)}h_{\mu\nu}\alpha \nn\\
&& +C_3\, \left[\bnabla_\mu\bnabla_\nu\(h\alpha\)+{2\Lambda\over (D-1)(D-2)}\bg_{\mu\nu}\(h\alpha\)\right]. \label{cubictransd}
\eea
There is one unfixed parameter $C_3$, which is just the linear transformation again with parameter $h\alpha$, so by re-defining the gauge parameter we can set $C_3=0$.

All the needed miracles occur only $D=4$, in other dimensions we are two short\footnote{Curiously, the prediction $D=4$ seems to agree with observations.}.   The zero derivative terms in \eqref{3ndordercond} are not canceled, except in $D=4$,
\be {\hat L}^{(0)}_{\mu\nu}{\delta{\cal L}^{(3)}_{}\over \delta h_{\mu\nu}}+{\hat L}^{(1)}_{\mu\nu}{\delta{\cal L}^{(2)}_{}\over \delta h_{\mu\nu}}={D-4\over 8(D-1)^2(D-2)}\Lambda^2\sqrt{-\bar g}\( h_{\mu\nu}^2-h^2\).
\ee

Note that the obstruction for $D\not=4$ vanishes in the decoupling limit, and curiously, it has the Fierz-Pauli structure.
This seems to imply that the partially massless theory as we have it can only exist in $D=4$ dimensions.  As has been noted, $D=4$ is special in that partially massless theories are conformally invariant there \cite{Deser:2004ji,Zinoviev:2006im}.   Also, this is the dimension in which the partially massless candidate has a $Z_2$ symmetry between the dynamical and background metrics.

\subsubsection{The cubic gauge transformation}

At quadratic order, the derivative part of the transformation had the form of a diffeomorphism, \ie we could write $\delta h_{\mu\nu}^{(0)}|_{\rm deriv} = \bnabla_\mu \xi_\nu^{(0)} + \bnabla_\nu \xi_\mu^{(0)}$, where $ \xi_\mu^{(0)} = \frac{1}{2} \bnabla_\mu \alpha$.  We can ask if the derivative part of the cubic transformation has the same property, that is, if we can write the non-linear transformation determined thus far as an expansion of
\be
\label{gderiv}
\delta h_{\mu\nu}|_{\rm deriv}= \nabla_\mu \xi_\nu+\nabla_\nu \xi_\mu \, ,
\ee
for some vector $\xi_\nu$.

To expand, we write the covariant derivative $\nabla$ in terms of the background covariant derivative $\bnabla$ and an associated connection $C^\lambda_{~\mu\nu}$
\be
\nabla_\mu \xi_\nu = \bnabla_\mu \xi_\nu-C^\lambda_{~\mu\nu} \xi_\lambda \, ,
\ee
\be
C^\lambda_{~\mu\nu} = \tfrac{1}{2}g^{\lambda \kappa}(\bnabla_\mu g_{\nu \kappa}+\bnabla_\nu g_{\mu \kappa}-\bnabla_\kappa g_{\mu \nu} )\, .
\ee
Expanding (\ref{gderiv}) in $h_{\mu\nu}$ gives
\be
\label{g1b}
\delta h_{\mu\nu}^{(1)} |_{\rm deriv} = \bnabla_\mu \xi_\nu^{(1)} + \bnabla_\nu \xi_\mu^{(1)}-2 C^{\lambda (1)}_{~\mu\nu} \xi_\lambda^{(0)}\, ,
\ee
for the transformation beyond leading order.

As determined by the analysis above, we have
\be
 \label{g1a}
 \delta h_{\mu\nu}^{(1)}|_{\rm deriv}  = \tfrac{1}{2}  h_{\lambda(\mu} \bnabla_{\nu)} \bnabla^\lambda \alpha +\tfrac{1}{2}\bnabla_\lambda h_{\mu\nu}\bnabla^\lambda \alpha -\tfrac{1}{2}\bnabla_{(\mu}h_{\nu)\lambda}  \bnabla^\lambda \alpha \, ,
\ee
for the derivative part of the cubic transformation.  Equating (\ref{g1a}) with (\ref{g1b}), we can solve:
\be
\xi_\mu^{(1)} =\tfrac{1}{4} h_{\mu\lambda}\bnabla^\lambda \alpha \, .
\ee
Thus, indeed, to cubic order the derivative part of the transformation has the form of a diffeomorphism with parameter
\be
\xi_\mu = \tfrac{1}{2}\bnabla_\mu \alpha +\tfrac{1}{4} h_{\mu\lambda}\bnabla^\lambda \alpha + {\cal O}(h^2) \, .
\ee

\section{Obstructions to Non-Linear PM Gravity\label{obstructionsection}}
\label{sec:Obs_Hints}

In this section we discuss obstructions to a general non-linear PM symmetry from several different angles.  In particular, we show that the candidate PM theory \eqref{LPM} does not have the required gauge symmetry to all orders.

\subsection{Anisotropic obstruction}

We now consider an extension of the mini-superspace analysis of Section \ref{frwsection1}, where we still restrict ourselves to an homogeneous ansatz but allow for anisotropies in the physical metric, so that the metric \eqref{FRW_metric} gets replaced by
\ba
\label{ani}
g\mn=-N(t)^2\d t^2+\sum_{i=1}^{D-1} a^i(t)^2 \d x^i{}^2\,,
\ea
while the reference metric $\bar g\mn$ remains dS.  We will show below that, while the isotropic ansatz had an evident gauge symmetry, this is not the case as soon as one allows for anisotropies.  This represents a first hint towards the absence of a PM symmetry in the fully general and non-linear case.

Let us focus on the case that $D=4$ so that our candidate action is given by
\be
{\cal S}={\mpl^{2} \over 2} \int \d^4x\ \sqrt{-g}\left(R[g]-2\lambda S_2(\sqrt{g^{-1}\bar g})\right) \, .
\ee
Using our anisotropic ansatz (\ref{ani}), and after integrations by parts to remove higher derivatives, the PM action in four dimensions is
\ba
{\cal S}=\frac{\mpl^2}{2}\int \d^4 x \left[
-\frac 2 N S^{(3)}_{ijk}a^i\dot a^j \dot a^k-2\lambda e^{Ht} S^{(2)}_{ij}a^i a^j-2 \lambda N e^{2 H t}S^{(1)}_{i}a^i
\right]\,.
\ea
We have defined the fully symmetric constant arrays $S^{(1)}_{i}$, $S^{(2)}_{ij}$, $S^{(3)}_{ijk}$ so that
\ba
&&  S^{(1)}_{i}a^i =a^1+a^2+a^3\\
&& S^{(2)}_{ij}a^ia^j =a^1a^2+a^2a^3+a^3a^1\\
&& S^{(3)}_{ijk}a^i\dot a^j \dot a^k =a^1\dot a^2\dot a^3+a^2\dot a^3\dot a^1+a^3\dot a^1\dot a^2\,.
\ea
We also define the shorthand $S^{(1)}\equiv S^{(1)}_{i}a^i $, $ S^{(2)}\equiv S^{(2)}_{ij}a^ia^j $, and $S^{(3)}\equiv S^{(3)}_{ijk}a^i\dot a^j \dot a^k $.

Similarly as in the isotropic case, the lapse is an auxiliary variable and we may solve for it,
\ba
N=\frac {e^{-H t}}{\sqrt{\lambda}}\sqrt{\frac{S^{(3)}}{S^{(1)}}}\,.
\ea
Plugging this constraint back into the Lagrangian, we get
\ba
\L=-\mpl^2e^{H t} \sqrt{\lambda}\left[2\sqrt{S^{(1)}S^{(3)}}+\sqrt{\lambda}S^{(2)}\right]\,.
\ea

We want to know whether this Lagrangian has a gauge symmetry, so we proceed with a Hamiltonian analysis.
The canonical momenta are
\ba\label{canonicalmomk}
\pi_i=\frac{\p \L}{\p \dot a^i}=-2\mpl^2 e^{H t}\sqrt{\lambda}\sqrt{\frac{S^{(1)}}{S^{(3)}}} S^{(3)}_{ijk}a^j \dot a^k\,.
\ea
Defining the `phase space metric' $G_{ij}$ as $G_{ij}=S^{(1)}S^{(3)}_{ijk}a^k$, and $G^{ij}$ its inverse, we see that the canonical variables satisfy $\pi_i G^{ij} \pi_j=4 \mpl^4 e^{2 H t} \lambda \,,$
so we have the primary constraint\footnote{We have square rooted both sides because this is the constraint that is regular in the isotropic case, and we have chosen the sign because the canonical momentum is negative \eqref{canonicalmomk}.}
\ba\label{primaryconst}
\mathcal{C}=\sqrt{\pi_i G^{ij} \pi_j}-2 \mpl^2 e^{H t}\sqrt{\lambda}\,.
\ea
The Hamiltonian is given by
\ba
\mathcal{H}= \pi_i \dot a^i-\L=\mpl^2 e^{ H t }\lambda S^{(2)}\,,
\ea
and does not depend on the canonical momenta.

If the primary constraint \eqref{primaryconst} is associated with a sole gauge symmetry, it should be the only constraint (in which case it is first class and generates the gauge symmetry).  Thus there should be no secondary constraint.
To look for a secondary constraint, following Dirac, we check the primary constraint's conservation in time,
\ba
\frac{\d \mathcal{C}}{\d t}=\frac{\p \mathcal{C}}{\p t}+\{{\cal C},{\cal H}\}=\frac{\p \mathcal{C}}{\p t}-\frac{\p \mathcal{C}}{\p \pi_i}\frac{\p \mathcal{H}}{\p a^i}=-2 \mpl^2e^{H t}\sqrt{\lambda}\(H + \sqrt{\lambda} \ \frac{G^{ij}\pi_j S^{(2)}_{ik}a^k}{\sqrt{\pi_\ell G^{\ell m}\pi_m}} \) \,.
\ea
In the generic anisotroic case, the right hand side does not vanish on the constraint surface, so we have a secondary constraint and hence not the single gauge invariance we are seeking.

In the isotropic case, $G^{ij}\pi_j S^{(2)}_{ik}a^k=\frac{\pi}{a}=-\sqrt{\pi_\ell G^{\ell m}\pi_m}\,$ (remembering $\pi<0$), and the right hand side vanishes when positing $\lambda=H^2$.  However we see now that this this gauge symmetry is lost as soon as we are dealing with a more general metric, even when $\lambda=H^2$. This is the first hint that the symmetry found in the previous section, either fully non-linearly in the minisuperspace, or at linear level does not propagate in full generality in the most naive version of the theory.

\subsection{Quartic obstruction from vectors in the decoupling limit}

We now move onto a second way to see that PM gravity fails to be realized at the non-linear level. For this, we push the analysis of the vector modes in Section \ref{decoupling} to quartic order in the fields.  We will see that the required gauge invariance of the vector modes cannot be realized at quartic order.

At quartic order in the vector field, the requirement \eqref{Abianchi} of gauge invariance for the decoupled vector mode Lagrangian reads
\be
\label{quarticAcond}
{\hat L}_{\mu}^{(0)}{\delta {\cal L}_{(4)}\over \delta A_\mu}+{\hat L}_{\mu}^{(1)}{\delta {\cal L}_{(3)}\over \delta A_\mu}+{\hat L}_{\mu}^{(2)}{\delta {\cal L}_{(2)}\over \delta A_\mu}=0\,.
\ee
The most general transformation ${\hat L}_{\mu}^{(2)}$ which is at most second order in derivatives and quadratic in the vector field is,
\bea {\hat L}_{\mu}^{(2)}\alpha=&& D_1\, A_\mu\, A^\nu\partial_\nu\alpha+D_2\, A^2\partial_\mu\alpha+D_3\,\partial_\mu A_\nu \, A^\nu \alpha \nn\\
&&+D_4\,\partial_\nu A_\mu\, A^\nu \alpha +D_5\, A_\mu\, (\partial A) \,\alpha .\label{vecqartans}
\eea

Plugging this into \eqref{quarticAcond}, and using the lower order relations from Section \ref{decoupling}, we find that no matter what the coefficients $D_1,\cdots,D_5$ of the transformation, and no matter what the choice of potential (\ie choice of coefficients $d_1,\cdots,d_5$), \eqref{quarticAcond} cannot be satisfied\footnote{In particular, if we look at the two tensor structures $\partial^\mu A^\nu \partial_\mu A^\rho \square \partial_\rho A_\nu$ and $\partial^\nu A^\mu \partial_\mu A^\rho \square \partial_\rho A_\nu$ coming from \eqref{quarticAcond}, we find their coefficients depend only on $d_1$ and $d_2$ in the following way
\be \( {3\over 8} - 2 d_1 - {3 \over 2}d_2\)\partial^\mu A^\nu \partial_\mu A^\rho \square \partial_\rho A_\nu+\( {5\over 8} - 2 d_1 - {3 \over 2}d_2 \)\partial^\nu A^\mu \partial_\mu A^\rho \square \partial_\rho A_\nu,\ee
and so both cannot be made to vanish simultaneously.} This remains true if we allow for up to 3rd derivatives in the ansatz \eqref{vecqartans} (as might be expected from counting derivatives in the decoupling limit.

We therefore conclude there is an obstruction to preserving only 4 DOF at the quartic level when considering a generic mass term in arbitrary dimensions. It is worth pointing out that this obstruction relies on the assumptions that a) the Einstein-Hilbert term receives no corrections, and b) that the gauge transformation is at most 3rd order in derivatives when acting on $A$.  It is possible that the obstruction could be relaxed by allowing more derivatives, however in that case we would also have more conditions to satisfy and it could be that the transformation would then involve an infinite number of derivatives and would thus likely be non-local\footnote{Note that this analysis of this section also serves to rule out the possibility that the flat-reference metric ``minimal model" (in the language of \cite{Hassan:2011vm}), for which the decoupling limit interactions vanish, has the higher interaction scale $\Lambda_{2}$ (in $D=4$).  This is because if the theory had this higher cutoff, the vector self-interactions in the $\Lambda_2$ decoupling limit would take the same form as they do in our analysis here, and since they cannot have a gauge symmetry non-linearly, they propagate more than two degrees of freedom non-linearly.  Then the decoupling limit theory (remembering the scalar kinetic term is still present) has more than the 5 degrees of freedom required of the ghost-free massive graviton.
This is consistent with the amplitude analysis of \cite{Schwartz:2003vj} which shows $\Lambda_3$ as the cutoff, which itself implies that the vector-scalar interactions in the decoupling limit of the minimal model cannot vanish.}.

\subsection{The full theory at quartic order in $D=4$}

Despite the encouraging evidence of section \ref{sec:Evidence} for the existence of a PM theory of gravity propagating only four degrees of freedom in four dimensions, the previous subsections already provided a few arguments obstructing the existence of a fully fledged PM theory of gravity.

Here, we report on our attempt to push the brute force calculation of Section \ref{fullcubicss} to fourth order in the fields in $D=4$ (we were obstructed already at cubic order when $D\not =4$).  We find that there is an obstruction, and the theory fails to be gauge invariant at this order.

At quartic order, the statement for gauge invariance \eqref{gaugeinv} reads
\bea &&{\hat L}^{(0)}_{\mu\nu}\left[-\half\sqrt{-g}\left.\(G^{\mu\nu}+\Lambda g^{\mu\nu}\)\right|_{(3)}+{\delta{\cal L}^{(4)}_{\rm m}\over \delta h_{\mu\nu}}\right]+{\hat L}^{(1)}_{\mu\nu}\left[-\half\sqrt{-g}\left.\(G^{\mu\nu}+\Lambda g^{\mu\nu}\)\right|_{(2)}+{\delta{\cal L}^{(3)}_{\rm m}\over \delta h_{\mu\nu}}\right] \nn\\
&&+{\hat L}^{(2)}_{\mu\nu}\left[-\half\sqrt{-g}\left.\(G^{\mu\nu}+\Lambda g^{\mu\nu}\)\right|_{(1)}+{\delta{\cal L}^{(2)}_{\rm m}\over \delta h_{\mu\nu}}\right]=0.\label{4ndordercond}
\eea
For ${\hat L}^{(2)}$, the most general form up to two derivatives has 72 terms (which we will not attempt to display here), with the same number of free coefficients.  In addition, we keep the quartic mass term general, which brings 5 free coefficients $d_1,\cdots,d_5$.  We plug this into \eqref{4ndordercond}, using the quadratic and cubic order results \eqref{zeroortrans} and \eqref{cubictransd} for the lowest order transformations, and we demand that \eqref{4ndordercond} holds.  We fix coefficients starting from the highest derivatives, fixing coefficients requiring the various terms of the form $h^2 \bnabla\bnabla\bnabla\bnabla h$ to vanish, then the terms $h \bnabla h \bnabla\bnabla\bnabla h$, then $\bnabla h \bnabla h\bnabla\bnabla h$.  All these four derivative terms can be eliminated.  Next we start on the two derivative terms, starting with those where both derivatives act on the same $h$, i.e. those of the form $h^2\bnabla\bnabla h$, which can also be eliminated.  Then we come to the two derivative terms of the form $h\bnabla h\bnabla h$, and it is here where an obstruction appears: these terms cannot be canceled without setting $\Lambda=0$.

In the process, however, the coefficients $d_1,\cdots,d_5$ in the quartic mass term in \eqref{startlag} are all uniquely determined,
\bea {\cal L}&&={\cal L}_{\rm EH}+{\cal L}_{\rm m} \nn\\
&&=\half \left[ \sqrt{-g}\(R[g]-2\Lambda\)-{m^2\over 4}\sqrt{-\bar g}\left( [h^2]-[h]^2-[h^3]+{5\over 4}[h^2][h]-{1\over 4}[h]^3 \right.\right. \nn\\
&&\left.\left. +{7\over 8}[ h^4]-{7\over 8} [ h^3] [ h] -{11\over 32}[ h^2 ]^2 +{3\over 8}[ h^2][ h]^2-{1\over 32}[ h]^4 +\cdots\right)\right]\nn\,,
\eea
with $m^2$ fixed to \eqref{m2value}. These mass term coefficients match precisely those coming from expanding the candidate theory \eqref{LPM} to quartic order (even though we started with the most generic non-derivative interactions). Nevertheless, it seems impossible to preserve a symmetry at quartic level and hence to ensure that the helicity-0 mode of massive gravity gets consistently removed, inhibiting the existence of a partially massless theory of gravity.  The only assumptions we made are that the kinetic term is that of Einstein-Hilbert, and the gauge symmetry has at most two derivatives.  This conclusion is mentioned without calculation in \cite{Zinoviev:2006im}, and reached as well with different methods in \cite{Deser:2013uy}.  In the case of $D=3$, new massive gravity \cite{Bergshoeff:2009hq} on a dS background also becomes a situation where the PM symmetry is present at linear level but does not extend to all orders \cite{Gabadadze:2012xv}.

\section{Outlook}
Unfortunately, despite the suggestive evidence presented in the first half of this paper, a brute force calculation demonstrates that, as formulated, the PM theory doesn't exist at the non-linear level.  Let us emphasize that while much of our analysis focused on a particular candidate theory (i.e., a specific choice of dRGT massive gravity), our ``no-go" result is more general.  In our perturbative analysis we placed no assumptions on the form of the non-derivative interactions and we found that no gauge symmetry can exist beyond cubic order, in any number of space-time dimensions.

There are still several ways in which the no-go result may be avoided.  In dRGT the kinetic term is that of Einstein-Hilbert, and even in our more general perturbative analysis we assume an Einstein-Hilbert kinetic term as well.  It could be that there are non-canonical or higher derivative kinetic terms, analogous to the Lovelock terms, which are still ghost-free and allow for partially massless interactions\footnote{Note, however, that we disagree with the conclusions of \cite{Folkerts:2011ev} regarding the existence of ghosts, see e.g. \cite{deRham:2011qq}.} \cite{Folkerts:2011ev,Hinterbichler:2013eza}.  Another possibility is that additional degrees of freedom are needed in order to have consistent interaction.  It could be that one of the bimetric or multi-metric versions of dRGT theory \cite{Hassan:2011zd,Hinterbichler:2012cn} has a gauge symmetry and contains a partially massless graviton.  Candidate bimetric theories are proposed in \cite{Hassan:2012gz,Hassan:2012rq} (see \cite{Hassan:2013pca,Deser:2013gpa} for arguments for and against).  However, the candidate PM bimetric theory can be related via a scaling limit to the single-metric PM theory, and does not have the PM symmetry \cite{Fasiello:2013woa}.

A theory which comes close to realizing a non-linear partially massless graviton is conformal gravity in $D=4$, whose action is the square of the Weyl tensor, as studied recently in \cite{Deser:2012qg}.  Around a dS solution, the theory propagates a massless graviton and a partially massless graviton \cite{Maldacena:2011mk}, but unfortunately precisely one is always a ghost.  Any attempt to truncate out this unwanted degree of freedom fails at quartic order in the fields \cite{Deser:2012qg}.  Nevertheless, the full theory has six degrees of freedom and is Weyl invariant, which acts as the partially massless symmetry \eqref{partialmasslesssym} at linear level.  Thus this theory serves as one example where it is possible to extend the PM gauge symmetry to all orders.

Finally, if PM is to be a phenomenologically viable theory of gravity, the issue of how the graviton couples to matter must be addressed.  Requiring that the matter sector maintains the partially massless symmetry will certainly put strong constraints on the form of the coupling.  For instance, at linear level the symmetry \eqref{partialmasslesssym} requires that the stress tensor satisfy
\be \(\bnabla_\mu\bnabla_\nu+{m^2\over D-2}\, \bg_{\mu\nu}\)T^{\mu\nu}=0.\ee
So the stress tensor does not have to be conserved, but its non-conservation is suppressed by Hubble scales.
Nonetheless, the possibility of addressing the old cosmological constant problem through the partially massless symmetry remains a compelling reason to continue the pursuit of PM gravity and its possible couplings.  While ultimately unsuccessful, the candidate theory of PM gravity studied here has enough promising features to serve as a starting point for future investigations.

\vskip.5cm

\bigskip
{\bf Acknowledgements}: We would like to thank Gregory Gabadadze and Mehrdad Mirbabayi for discussions.  CdR and AJT would like to thank Luana Kalinda Tolley de Rham for important feedback during the writing of this paper. AJT was supported in part by the Department of Energy under grant DE-FG02-12ER41810.  Research at Perimeter Institute is supported by the Government of Canada through Industry Canada and by the Province of Ontario through the Ministry of Economic Development and Innovation.  This work was made possible in part through the support of a grant from the John Templeton Foundation. The opinions expressed in this publication are those of the authors and do not necessarily reflect the views of the John Templeton Foundation (KH).  RAR was supported by NASA contract NNX10AH14G and US Department of Energy grant DE- FG02-11ER41743.

\bibliographystyle{utphys}
\addcontentsline{toc}{section}{References}
\bibliography{PMv10}

\providecommand{\href}[2]{#2}\begingroup\raggedright\begin{thebibliography}{10}

\bibitem{Hinterbichler:2011tt}
K.~Hinterbichler, ``{Theoretical Aspects of Massive Gravity},''
  \href{http://dx.doi.org/10.1103/RevModPhys.84.671}{{\em Rev.Mod.Phys.}
  {\bfseries 84} (2012) 671--710},
\href{http://arxiv.org/abs/1105.3735}{{\ttfamily arXiv:1105.3735 [hep-th]}}.

\bibitem{Deser:1983mm}
S.~Deser and R.~I. Nepomechie, ``{Gauge Invariance Versus Masslessness In De
  Sitter Space},''
\href{http://dx.doi.org/10.1016/0003-4916(84)90156-8}{{\em Ann. Phys.}
  {\bfseries 154} (1984) 396}.

\bibitem{Higuchi:1986py}
A.~Higuchi, ``{Forbidden mass range for spin-2 field theory in de Sitter space
  time},''
\href{http://dx.doi.org/10.1016/0550-3213(87)90691-2}{{\em Nucl.Phys.}
  {\bfseries B282} (1987) 397}.

\bibitem{Deser:2001pe}
S.~Deser and A.~Waldron, ``{Gauge invariances and phases of massive higher
  spins in (A)dS},''
  \href{http://dx.doi.org/10.1103/PhysRevLett.87.031601}{{\em Phys.Rev.Lett.}
  {\bfseries 87} (2001) 031601},
\href{http://arxiv.org/abs/hep-th/0102166}{{\ttfamily arXiv:hep-th/0102166
  [hep-th]}}.

\bibitem{Deser:2001us}
S.~Deser and A.~Waldron, ``{Partial masslessness of higher spins in (A)dS},''
  \href{http://dx.doi.org/10.1016/S0550-3213(01)00212-7}{{\em Nucl.Phys.}
  {\bfseries B607} (2001) 577--604},
\href{http://arxiv.org/abs/hep-th/0103198}{{\ttfamily arXiv:hep-th/0103198
  [hep-th]}}.

\bibitem{Deser:2001wx}
S.~Deser and A.~Waldron, ``{Stability of massive cosmological gravitons},''
  \href{http://dx.doi.org/10.1016/S0370-2693(01)00523-8}{{\em Phys.Lett.}
  {\bfseries B508} (2001) 347--353},
\href{http://arxiv.org/abs/hep-th/0103255}{{\ttfamily arXiv:hep-th/0103255
  [hep-th]}}.

\bibitem{Deser:2001xr}
S.~Deser and A.~Waldron, ``{Null propagation of partially massless higher spins
  in (A)dS and cosmological constant speculations},''
  \href{http://dx.doi.org/10.1016/S0370-2693(01)00756-0}{{\em Phys.Lett.}
  {\bfseries B513} (2001) 137--141},
\href{http://arxiv.org/abs/hep-th/0105181}{{\ttfamily arXiv:hep-th/0105181
  [hep-th]}}.

\bibitem{Deser:2004ji}
S.~Deser and A.~Waldron, ``{Conformal invariance of partially massless higher
  spins},'' \href{http://dx.doi.org/10.1016/j.physletb.2004.10.007}{{\em
  Phys.Lett.} {\bfseries B603} (2004) 30},
\href{http://arxiv.org/abs/hep-th/0408155}{{\ttfamily arXiv:hep-th/0408155
  [hep-th]}}.

\bibitem{Zinoviev:2001dt}
Y.~Zinoviev, ``{On massive high spin particles in AdS},''
\href{http://arxiv.org/abs/hep-th/0108192}{{\ttfamily arXiv:hep-th/0108192
  [hep-th]}}.

\bibitem{ArkaniHamed:2002sp}
N.~Arkani-Hamed, H.~Georgi, and M.~D. Schwartz, ``{Effective field theory for
  massive gravitons and gravity in theory space},''
  \href{http://dx.doi.org/10.1016/S0003-4916(03)00068-X}{{\em Annals Phys.}
  {\bfseries 305} (2003) 96--118},
\href{http://arxiv.org/abs/hep-th/0210184}{{\ttfamily arXiv:hep-th/0210184
  [hep-th]}}.

\bibitem{deRham:2012ew}
C.~de~Rham, G.~Gabadadze, L.~Heisenberg, and D.~Pirtskhalava,
  ``{Non-Renormalization and Naturalness in a Class of Scalar-Tensor
  Theories},''
\href{http://arxiv.org/abs/1212.4128}{{\ttfamily arXiv:1212.4128 [hep-th]}}.

\bibitem{ArkaniHamed:2002fu}
N.~Arkani-Hamed, S.~Dimopoulos, G.~Dvali, and G.~Gabadadze, ``{Nonlocal
  modification of gravity and the cosmological constant problem},''
\href{http://arxiv.org/abs/hep-th/0209227}{{\ttfamily arXiv:hep-th/0209227
  [hep-th]}}.

\bibitem{Dvali:2007kt}
G.~Dvali, S.~Hofmann, and J.~Khoury, ``{Degravitation of the cosmological
  constant and graviton width},''
  \href{http://dx.doi.org/10.1103/PhysRevD.76.084006}{{\em Phys.Rev.}
  {\bfseries D76} (2007) 084006},
\href{http://arxiv.org/abs/hep-th/0703027}{{\ttfamily arXiv:hep-th/0703027
  [HEP-TH]}}.

\bibitem{vanDam:1970vg}
H.~van Dam and M.~J.~G. Veltman, ``{Massive and massless Yang-Mills and
  gravitational fields},''
\href{http://dx.doi.org/10.1016/0550-3213(70)90416-5}{{\em Nucl. Phys.}
  {\bfseries B22} (1970) 397--411}.

\bibitem{Zakharov:1970cC}
V.~Zakharov, ``{Linearized gravitation theory and the graviton mass},''
{\em JETP Lett.} {\bfseries 12} (1970) 312.

\bibitem{Vainshtein:1972sx}
A.~I. Vainshtein, ``{To the problem of nonvanishing gravitation mass},''
\href{http://dx.doi.org/10.1016/0370-2693(72)90147-5}{{\em Phys. Lett.}
  {\bfseries B39} (1972) 393--394}.

\bibitem{deRham:2011pt}
C.~de~Rham, G.~Gabadadze, and A.~J. Tolley, ``{Comments on
  (super)luminality},''
\href{http://arxiv.org/abs/1107.0710}{{\ttfamily arXiv:1107.0710 [hep-th]}}.

\bibitem{Burrage:2011cr}
C.~Burrage, C.~de~Rham, L.~Heisenberg, and A.~J. Tolley, ``{Chronology
  Protection in Galileon Models and Massive Gravity},''
  \href{http://dx.doi.org/10.1088/1475-7516/2012/07/004}{{\em JCAP} {\bfseries
  1207} (2012) 004},
\href{http://arxiv.org/abs/1111.5549}{{\ttfamily arXiv:1111.5549 [hep-th]}}.

\bibitem{Izumi:2013poa}
K.~Izumi and Y.~C. Ong, ``{An Analysis of Characteristics in Non-Linear Massive
  Gravity},''
\href{http://arxiv.org/abs/1304.0211}{{\ttfamily arXiv:1304.0211 [hep-th]}}.

\bibitem{Deser:2013eua}
S.~Deser, K.~Izumi, Y.~Ong, and A.~Waldron, ``{Massive Gravity Acausality
  Redux},''
\href{http://arxiv.org/abs/1306.5457}{{\ttfamily arXiv:1306.5457 [hep-th]}}.

\bibitem{Deser:2013uy}
S.~Deser, M.~Sandora, and A.~Waldron, ``{Nonlinear Partially Massless from
  Massive Gravity?},''
\href{http://arxiv.org/abs/1301.5621}{{\ttfamily arXiv:1301.5621 [hep-th]}}.

\bibitem{deRham:2010iK}
C.~de~Rham and G.~Gabadadze, ``{Generalization of the Fierz-Pauli Action},''
  \href{http://dx.doi.org/10.1103/PhysRevD.82.044020}{{\em Phys. Rev.}
  {\bfseries D82} (2010) 044020},
\href{http://arxiv.org/abs/1007.0443}{{\ttfamily arXiv:1007.0443 [hep-th]}}.

\bibitem{deRham:2010kj}
C.~de~Rham, G.~Gabadadze, and A.~J. Tolley, ``{Resummation of Massive
  Gravity},'' \href{http://dx.doi.org/10.1103/PhysRevLett.106.231101}{{\em
  Phys.Rev.Lett.} {\bfseries 106} (2011) 231101},
\href{http://arxiv.org/abs/1011.1232}{{\ttfamily arXiv:1011.1232 [hep-th]}}.

\bibitem{Gumrukcuoglu:2011zh}
A.~E. Gumrukcuoglu, C.~Lin, and S.~Mukohyama, ``{Cosmological perturbations of
  self-accelerating universe in nonlinear massive gravity},''
  \href{http://dx.doi.org/10.1088/1475-7516/2012/03/006}{{\em JCAP} {\bfseries
  1203} (2012) 006},
\href{http://arxiv.org/abs/1111.4107}{{\ttfamily arXiv:1111.4107 [hep-th]}}.

\bibitem{D'Amico:2012pi}
G.~D'Amico, ``{Cosmology and perturbations in massive gravity},''
  \href{http://dx.doi.org/10.1103/PhysRevD.86.124019}{{\em Phys.Rev.}
  {\bfseries D86} (2012) 124019},
\href{http://arxiv.org/abs/1206.3617}{{\ttfamily arXiv:1206.3617 [hep-th]}}.

\bibitem{Wyman:2012iw}
M.~Wyman, W.~Hu, and P.~Gratia, ``{Self-accelerating Massive Gravity: Time for
  Field Fluctuations},''
  \href{http://dx.doi.org/10.1103/PhysRevD.87.084046}{{\em Phys.Rev.}
  {\bfseries D87} (2013) 084046},
\href{http://arxiv.org/abs/1211.4576}{{\ttfamily arXiv:1211.4576 [hep-th]}}.

\bibitem{Khosravi:2013axa}
N.~Khosravi, G.~Niz, K.~Koyama, and G.~Tasinato, ``{Stability of the
  Self-accelerating Universe in Massive Gravity},''
\href{http://arxiv.org/abs/1305.4950}{{\ttfamily arXiv:1305.4950 [hep-th]}}.

\bibitem{Rubakov:2008nh}
V.~Rubakov and P.~Tinyakov, ``{Infrared-modified gravities and massive
  gravitons},'' \href{http://dx.doi.org/10.1070/PU2008v051n08ABEH006600}{{\em
  Phys.Usp.} {\bfseries 51} (2008) 759--792},
\href{http://arxiv.org/abs/0802.4379}{{\ttfamily arXiv:0802.4379 [hep-th]}}.

\bibitem{deRham:2012kf}
C.~de~Rham and S.~Renaux-Petel, ``{Massive Gravity on de Sitter and Unique
  Candidate for Partially Massless Gravity},''
\href{http://arxiv.org/abs/1206.3482}{{\ttfamily arXiv:1206.3482 [hep-th]}}.

\bibitem{Zinoviev:2006im}
Y.~Zinoviev, ``{On massive spin 2 interactions},''
  \href{http://dx.doi.org/10.1016/j.nuclphysb.2007.02.005}{{\em Nucl.Phys.}
  {\bfseries B770} (2007) 83--106},
\href{http://arxiv.org/abs/hep-th/0609170}{{\ttfamily arXiv:hep-th/0609170
  [hep-th]}}.

\bibitem{Joung:2012hz}
E.~Joung, L.~Lopez, and M.~Taronna, ``{Generating functions of
  (partially-)massless higher-spin cubic interactions},''
\href{http://arxiv.org/abs/1211.5912}{{\ttfamily arXiv:1211.5912 [hep-th]}}.

\bibitem{Breitenlohner:1982bm}
P.~Breitenlohner and D.~Z. Freedman, ``{Positive Energy in anti-De Sitter
  Backgrounds and Gauged Extended Supergravity},''
\href{http://dx.doi.org/10.1016/0370-2693(82)90643-8}{{\em Phys.Lett.}
  {\bfseries B115} (1982) 197}.

\bibitem{Deser:2006zx}
S.~Deser and A.~Waldron, ``{Partially Massless Spin 2 Electrodynamics},''
  \href{http://dx.doi.org/10.1103/PhysRevD.74.084036}{{\em Phys.Rev.}
  {\bfseries D74} (2006) 084036},
\href{http://arxiv.org/abs/hep-th/0609113}{{\ttfamily arXiv:hep-th/0609113
  [hep-th]}}.

\bibitem{Deser:2013xb}
S.~Deser and A.~Waldron, ``{PM = EM: Partially Massless Duality Invariance},''
\href{http://arxiv.org/abs/1301.2238}{{\ttfamily arXiv:1301.2238 [hep-th]}}.

\bibitem{Boulware:1973my}
D.~G. Boulware and S.~Deser, ``{Can gravitation have a finite range?},''
\href{http://dx.doi.org/10.1103/PhysRevD.6.3368}{{\em Phys. Rev.} {\bfseries
  D6} (1972) 3368--3382}.

\bibitem{Hassan:2011hr}
S.~Hassan and R.~A. Rosen, ``{Resolving the Ghost Problem in non-Linear Massive
  Gravity},'' \href{http://dx.doi.org/10.1103/PhysRevLett.108.041101}{{\em
  Phys.Rev.Lett.} {\bfseries 108} (2012) 041101},
\href{http://arxiv.org/abs/1106.3344}{{\ttfamily arXiv:1106.3344 [hep-th]}}.

\bibitem{deRham:2011rn}
C.~de~Rham, G.~Gabadadze, and A.~Tolley, ``{Ghost free Massive Gravity in the
  St\'uckelberg language},''
\href{http://arxiv.org/abs/1107.3820}{{\ttfamily arXiv:1107.3820 [hep-th]}}.

\bibitem{deRham:2011qq}
C.~de~Rham, G.~Gabadadze, and A.~J. Tolley, ``{Helicity Decomposition of
  Ghost-free Massive Gravity},''
  \href{http://dx.doi.org/10.1007/JHEP11(2011)093}{{\em JHEP} {\bfseries 1111}
  (2011) 093},
\href{http://arxiv.org/abs/1108.4521}{{\ttfamily arXiv:1108.4521 [hep-th]}}.

\bibitem{Hassan:2011tf}
S.~Hassan, R.~A. Rosen, and A.~Schmidt-May, ``{Ghost-free Massive Gravity with
  a General Reference Metric},''
  \href{http://dx.doi.org/10.1007/JHEP02(2012)026}{{\em JHEP} {\bfseries 1202}
  (2012) 026},
\href{http://arxiv.org/abs/1109.3230}{{\ttfamily arXiv:1109.3230 [hep-th]}}.

\bibitem{Hassan:2011ea}
S.~Hassan and R.~A. Rosen, ``{Confirmation of the Secondary Constraint and
  Absence of Ghost in Massive Gravity and Bimetric Gravity},''
  \href{http://dx.doi.org/10.1007/JHEP04(2012)123}{{\em JHEP} {\bfseries 1204}
  (2012) 123},
\href{http://arxiv.org/abs/1111.2070}{{\ttfamily arXiv:1111.2070 [hep-th]}}.

\bibitem{Mirbabayi:2011aa}
M.~Mirbabayi, ``{A Proof Of Ghost Freedom In de Rham-Gabadadze-Tolley Massive
  Gravity},''
\href{http://arxiv.org/abs/1112.1435}{{\ttfamily arXiv:1112.1435 [hep-th]}}.

\bibitem{Golovnev:2011aa}
A.~Golovnev, ``{On the Hamiltonian analysis of non-linear massive gravity},''
  \href{http://dx.doi.org/10.1016/j.physletb.2011.12.064}{{\em Phys.Lett.}
  {\bfseries B707} (2012) 404--408},
\href{http://arxiv.org/abs/1112.2134}{{\ttfamily arXiv:1112.2134 [gr-qc]}}.

\bibitem{Hassan:2012qv}
S.~Hassan, A.~Schmidt-May, and M.~von Strauss, ``{Proof of Consistency of
  Nonlinear Massive Gravity in the St\'uckelberg Formulation},''
  \href{http://dx.doi.org/10.1016/j.physletb.2012.07.018}{{\em Phys.Lett.}
  {\bfseries B715} (2012) 335--339},
\href{http://arxiv.org/abs/1203.5283}{{\ttfamily arXiv:1203.5283 [hep-th]}}.

\bibitem{Hinterbichler:2012cn}
K.~Hinterbichler and R.~A. Rosen, ``{Interacting Spin-2 Fields},''
  \href{http://dx.doi.org/10.1007/JHEP07(2012)047}{{\em JHEP} {\bfseries 1207}
  (2012) 047},
\href{http://arxiv.org/abs/1203.5783}{{\ttfamily arXiv:1203.5783 [hep-th]}}.

\bibitem{Kluson:2012wf}
J.~Kluson, ``{Non-Linear Massive Gravity with Additional Primary Constraint and
  Absence of Ghosts},''
  \href{http://dx.doi.org/10.1103/PhysRevD.86.044024}{{\em Phys.Rev.}
  {\bfseries D86} (2012) 044024},
\href{http://arxiv.org/abs/1204.2957}{{\ttfamily arXiv:1204.2957 [hep-th]}}.

\bibitem{Hassan:2011zd}
S.~Hassan and R.~A. Rosen, ``{Bimetric Gravity from Ghost-free Massive
  Gravity},'' \href{http://dx.doi.org/10.1007/JHEP02(2012)126}{{\em JHEP}
  {\bfseries 1202} (2012) 126},
\href{http://arxiv.org/abs/1109.3515}{{\ttfamily arXiv:1109.3515 [hep-th]}}.

\bibitem{Hassan:2012gz}
S.~Hassan, A.~Schmidt-May, and M.~von Strauss, ``{On Partially Massless
  Bimetric Gravity},''
\href{http://arxiv.org/abs/1208.1797}{{\ttfamily arXiv:1208.1797 [hep-th]}}.

\bibitem{Hassan:2012rq}
S.~Hassan, A.~Schmidt-May, and M.~von Strauss, ``{Bimetric Theory and Partial
  Masslessness with Lanczos-Lovelock Terms in Arbitrary Dimensions},''
\href{http://arxiv.org/abs/1212.4525}{{\ttfamily arXiv:1212.4525 [hep-th]}}.

\bibitem{Fasiello:2013woa}
M.~Fasiello and A.~J. Tolley, ``{Cosmological Stability Bound in Massive
  Gravity and Bigravity},''
\href{http://arxiv.org/abs/1308.1647}{{\ttfamily arXiv:1308.1647 [hep-th]}}.

\bibitem{Skvortsov:2006at}
E.~Skvortsov and M.~Vasiliev, ``{Geometric formulation for partially massless
  fields},'' \href{http://dx.doi.org/10.1016/j.nuclphysb.2006.06.019}{{\em
  Nucl.Phys.} {\bfseries B756} (2006) 117--147},
\href{http://arxiv.org/abs/hep-th/0601095}{{\ttfamily arXiv:hep-th/0601095
  [hep-th]}}.

\bibitem{D'Amico:2011jj}
G.~D'Amico, C.~de~Rham, S.~Dubovsky, G.~Gabadadze, D.~Pirtskhalava, {\em
  et~al.}, ``{Massive Cosmologies},''
  \href{http://dx.doi.org/10.1103/PhysRevD.84.124046}{{\em Phys.Rev.}
  {\bfseries D84} (2011) 124046},
\href{http://arxiv.org/abs/1108.5231}{{\ttfamily arXiv:1108.5231 [hep-th]}}.

\bibitem{Fasiello:2012rw}
M.~Fasiello and A.~J. Tolley, ``{Cosmological perturbations in Massive Gravity
  and the Higuchi bound},''
\href{http://arxiv.org/abs/1206.3852}{{\ttfamily arXiv:1206.3852 [hep-th]}}.

\bibitem{Hassan:2011vm}
S.~Hassan and R.~A. Rosen, ``{On Non-Linear Actions for Massive Gravity},''
  \href{http://dx.doi.org/10.1007/JHEP07(2011)009}{{\em JHEP} {\bfseries 1107}
  (2011) 009},
\href{http://arxiv.org/abs/1103.6055}{{\ttfamily arXiv:1103.6055 [hep-th]}}.

\bibitem{Schwartz:2003vj}
M.~D. Schwartz, ``{Constructing gravitational dimensions},''
  \href{http://dx.doi.org/10.1103/PhysRevD.68.024029}{{\em Phys.Rev.}
  {\bfseries D68} (2003) 024029},
\href{http://arxiv.org/abs/hep-th/0303114}{{\ttfamily arXiv:hep-th/0303114
  [hep-th]}}.

\bibitem{Bergshoeff:2009hq}
E.~A. Bergshoeff, O.~Hohm, and P.~K. Townsend, ``{Massive Gravity in Three
  Dimensions},'' \href{http://dx.doi.org/10.1103/PhysRevLett.102.201301}{{\em
  Phys.Rev.Lett.} {\bfseries 102} (2009) 201301},
\href{http://arxiv.org/abs/0901.1766}{{\ttfamily arXiv:0901.1766 [hep-th]}}.

\bibitem{Gabadadze:2012xv}
G.~Gabadadze, G.~Giribet, and A.~Iglesias, ``{New Massive Gravity on de Sitter
  Space and Black Holes at the Special Point},''
\href{http://arxiv.org/abs/1212.6279}{{\ttfamily arXiv:1212.6279 [hep-th]}}.

\bibitem{Folkerts:2011ev}
S.~Folkerts, A.~Pritzel, and N.~Wintergerst, ``{On ghosts in theories of
  self-interacting massive spin-2 particles},''
\href{http://arxiv.org/abs/1107.3157}{{\ttfamily arXiv:1107.3157 [hep-th]}}.

\bibitem{Hinterbichler:2013eza}
K.~Hinterbichler, ``{Ghost-Free Derivative Interactions for a Massive
  Graviton},''
\href{http://arxiv.org/abs/1305.7227}{{\ttfamily arXiv:1305.7227 [hep-th]}}.

\bibitem{Hassan:2013pca}
S.~Hassan, A.~Schmidt-May, and M.~von Strauss, ``{Higher Derivative Gravity and
  Conformal Gravity From Bimetric and Partially Massless Bimetric Theory},''
\href{http://arxiv.org/abs/1303.6940}{{\ttfamily arXiv:1303.6940 [hep-th]}}.

\bibitem{Deser:2013gpa}
S.~Deser, M.~Sandora, and A.~Waldron, ``{No consistent bimetric gravity?},''
\href{http://arxiv.org/abs/1306.0647}{{\ttfamily arXiv:1306.0647 [hep-th]}}.

\bibitem{Deser:2012qg}
S.~Deser, E.~Joung, and A.~Waldron, ``{Gravitational and self-coupling of
  partially massless spin 2},''
  \href{http://dx.doi.org/10.1103/PhysRevD.86.104004}{{\em Phys.Rev.}
  {\bfseries D86} (2012) 104004},
\href{http://arxiv.org/abs/1208.1307}{{\ttfamily arXiv:1208.1307 [hep-th]}}.

\bibitem{Maldacena:2011mk}
J.~Maldacena, ``{Einstein Gravity from Conformal Gravity},''
\href{http://arxiv.org/abs/1105.5632}{{\ttfamily arXiv:1105.5632 [hep-th]}}.

\end{thebibliography}\endgroup

\end{document}